\documentclass[fleqn,usenatbib]{mnras}

\usepackage{newtxtext,newtxmath}
\usepackage[T1]{fontenc}

\DeclareRobustCommand{\VAN}[3]{#2}
\let\VANthebibliography\thebibliography
\def\thebibliography{\DeclareRobustCommand{\VAN}[3]{##3}\VANthebibliography}

\usepackage{graphicx}	%
\usepackage{amsmath}	%
\usepackage{graphicx}%
\DeclareMathOperator*{\argminB}{argmin}
\title[Scalable normalizing flows VLBI imaging]{
  Flow-based neural networks for scalable high resolution imaging of VLBI
  observations
}

\author[L. E. Kats et al.]{
L.E. Kats,$^{1}$\thanks{E-mail: l.kats@lebedev.ru}
A.B. Goldin,$^{2}$\thanks{E-mail: alexey.goldin@gmail.com}
A.S. Andrianov.$^{1}$\thanks{E-mail: andrian@asc.rssi.ru}
\\
$^{1}$Astro Space Center, Lebedev Physical Institute, Russian Academy of Sciences, Profsoyuznaya str. 84/32, Moscow, 117997, Russia\\
$^{2}$Teza Technologies LLC, 301 Congress Ave, 22nd Floor, Austin, TX 78701, USA
}

\date{Accepted 2026 September 25. Received 2026 September 25; in original form 2025 December 31}

\pubyear{2026}

\begin{document}
\label{firstpage}
\pagerange{\pageref{firstpage}--\pageref{lastpage}}
\maketitle

\begin{abstract}
Bayesian sampling methods in very-long-baseline interferometry enable reliable
imaging of astronomical sources with a record high angular resolution. However,
even after the transition from classical MCMC methods to flow-based neural
networks, they remain computationally expensive as the number of recovered
features grows. We used WaveletFlow neural network architecture to increase the
number of pixels in the recovered image distribution, addressing the scaling
issues of previous models. We applied the algorithm to both model data and
archive RadioAstron project observations of M87* and  ALMA telescope
observations of HL Tau. While yielding comparable fidelity of the individual
images, the algorithm provides more statistical information compared to CLEAN,
which may be crucial for future research on compact astronomical objects such
as active galactic nuclei.
\end{abstract}


\section{Introduction}

VLBI imaging since its introduction has been one of the main ways
to explore distant compact objects like AGNs.
However, an interferometer does not produce direct measurements of the spatial
luminosity distribution of an object. Instead, it samples a sparse and
incomplete set of Fourier components of the image. As a result, the imaging
problem is inverse and ill-posed, thus having infinitely many solutions.

There are many algorithms aimed to solve this problem, but because of the
problem being ill-posed, at least some prior information about the observed
object is usually assumed either explicitly or implicitly. Let us consider
for example the most well-known and well-understood imaging algorithms: CLEAN
\citep{Hogbom1974} and maximum entropy method \citep[MEM,][]{Narayan1986}. The
prior information used in CLEAN is that the recovered image may be expressed as
a sum of a relatively small number of shifted basis functions, which is
considered somewhat equivalent to l1-norm regularization of the problem
\citep{Marsh1987}. In the original work, delta functions were used as a basis
functions, but other options, including Gaussians \citep{Cornwell2008} and even
wavelets \citep{Muller2023} do exist. In MEM, on the other hand,
the problem is regularized with some form of entropy which results in the
algorithm being biased towards the solution that produced less commitment.

Any new method capable of using different types of prior information resulting
in new solutions is of great value. Such solutions are valuable because they
help distinguish real image features encoded in the observational data from
artefacts introduced by the imaging algorithm
\citep[see for example ][]{EHT2019,Fuentes2023}.
However, while classical
methods above can choose some meaningful and interpretable solutions,
extrapolating missing Fourier coefficients, significant work has been made
towards the algorithms able to accurately model the data measurement errors,
leading us to Bayesian algorithms. They can be roughly divided into two groups:
'point estimate' and sampling methods. Examples of the former usually can be
classified as regularized maximum likelihood methods
\citep[see for example ][]{Chael2018}, they produce single 'most likely' image as
a result. However, while they improve upon classical approaches, sampling methods
result in a sample of images, fully projecting measurement errors into the
solution space by fitting the posterior image distribution.

The classical approach to the sampling problem is to use some form of Markov
chain Monte Carlo (MCMC) method. Various algorithms have been proposed, see
\citet{Tiede2025}. However, MCMC methods often suffer from the curse of
dimensionality: naive approaches scale poorly as the number of recovered
parameters increases, while some approaches require some simplifying
assumptions and are unable to express arbitrary target distributions.
This work focuses on a family of algorithms based on normalizing flow
\citep{Dinh2014} neural network architectures, which are theoretically capable
of approximating arbitrary probability distributions \citep{Draxler2024}.

This work follows the framework proposed in \citet{Sun2020}. While the authors
of \citet{Sun2020} consider mostly imaging directly from closure quantities
and some relatively low-dimensional parametrizations of the resulting image
(e.g. ring parameters or images with size of at most 32x32 pixels), this work
focuses mainly on the 'convex' formulation of the imaging problem where
visibilities are used to find the spectral flux density in each pixel of the
target image and scaling the model to obtain images with as good resolution in
terms of total number of pixels as possible. We use the WaveletFlow
\citep{Yu2020} neural network architecture to deal with increased problem
dimensionality through adapting the multiscale approach. The algorithm is
tested both on model observations and calibrated archive data of RadioAstron
project observations of M87* and ALMA telescope observations of HL Tau. The
resulting images are compared with baselines produced by CLEAN.

\section{Method}

Let us describe the framework described in \citet{Sun2020} with regard to the
problem of imaging. Consider the inverse problem in form of the following
equation:
\begin{eqnarray}
  \mathbf{y} = \mathbf{F}(\mathbf{x}) + \mathbf{\delta y},
\end{eqnarray}
where $\mathbf{y}$ is the random vector of observations, that can be viewed as
a noised version of the ground truth measurement vector $\mathbf{y_0}$:
\begin{eqnarray}
  \mathbf{y} = \mathbf{y_0} + \mathbf{\delta y},
\end{eqnarray}
which remains unknown as well as the corresponding solution:
\begin{eqnarray}
  \mathbf{y_0} = \mathbf{F}(\mathbf{x}).
\end{eqnarray}

However, the probability density function of the noise $\mathbf{\delta y}$ is
known:
\begin{eqnarray}
  p(\mathbf{y} | \mathbf{x}) \equiv p(\mathbf{y} | \mathbf{y_0})
  \equiv p(\mathbf{\delta y}).
\end{eqnarray}
Also, some prior probability density function of possible solutions is assumed:
\begin{eqnarray}
  \mathbf{x} \sim p(\mathbf{x}).
\end{eqnarray}

Then we use the reparametrization trick similar to one used in VAEs
\cite{Kingma2013}:
\begin{eqnarray}
  \mathbf{x} &=& \mathbf{G}_{\mathbf{\theta}}(\mathbf{z}), \\
  \mathbf{z} &\sim& \mathcal{N}(0,\, 1) \nonumber
\end{eqnarray}
where $z$ is a vector of independent and identically distributed (iid)
real numbers with standard normal distribution,
$G$ is a neural network, parameterized with its weights $\theta$ so the PDF
of the model output $q_{\theta}$ can be expressed using the change of variable
formula:
\begin{eqnarray}
  q_{\mathbf{\theta}}(\mathbf{x}) =
  \pi(\mathbf{z})
  \left| \det
    \frac{\partial \mathbf{G}_{\theta}(\mathbf{z})}{\partial \mathbf{z}}
  \right|^{-1},
  \label{eq:dens}
\end{eqnarray}
where $\pi(z)$ is the PDF of standard normal distribution. One can assume that
the Jacobian matrix of the model is not degenerate which is assured by using
flow-based models and discussed in Section~\ref{sec:arch}.

The target distribution $q_{\theta}$ should be close to the posterior
distribution of solutions $p(\mathbf{x} | \mathbf{y})$, conditioned on the
observation vector. The popular choice is to make these distributions close in
the sense of Kullback–Leibler divergence. Combined with reparametrization
this is equivalent \citep{Sun2020} to the following optimization problem:
\begin{eqnarray}
  \begin{aligned}
  \mathbf{\theta} \longleftarrow & \argminB_{\mathbf{\theta}}
    \mathbb{E}_{\mathbf{z} \sim \mathcal{N}(0,\, 1)}\left[
      - \log p\left(\mathbf{y} | \mathbf{x}\right)
      - \log p\left(\mathbf{x}\right) + \right. \\ & \left.
      + \log \pi(\mathbf{z})
      - \log \left|
        \det \frac{
          \partial \mathbf{x}
        }{\partial \mathbf{z}}
      \right|
    \right], \\
  \mathbf{x} = & \mathbf{G}_{\mathbf{\theta}}(\mathbf{z}).
  \end{aligned}
  \label{eq:kldiv}
\end{eqnarray}

Let us consider errors in visibilities that consist of multiplicative
component, dominated by systematic calibration errors, and additive component,
caused by thermal noise, so one can write the measurement operator in the
following way:
\begin{equation}
  \tilde{V}_{a,b} (t) = g_a (t) g_b^* (t)
  \left( V_{a,b} (t) + \epsilon_{a,b} (t) \right),
  \label{eq:vis_errors}
\end{equation}
where $\tilde{V}_{a,b}$ are the measured correlations, $g_a$, $g_b$ are
station-specific relatively slow-changing complex functions of the time of the
scan, $\epsilon_{a,b}$ are the baseline-dependent thermal noises and $V_{a,b}$
are the true Fourier modes of the observed image. In this work we assume the
data are calibrated perfectly so the only considered source of error is
additive error noise (See Section~\ref{sec:disc} for the rationale). This is
referred to as the convex case in \citet{Sun2020}. The thermal noises
$\epsilon_{a,b}(t)$ are independent complex random variables with symmetric
Gaussian distribution with known variance $\sigma^2_{a,b}(t)$ of each
component, which give us the following expression:
\begin{equation}
\begin{aligned}
  \log p\left(\mathbf{y} | \mathbf{x}\right) &= -\sum_{a,b,j}
  \frac{\left|\tilde{V}_{a,b}(t_j) - V_{a,b} (t_j)\right|^2}
  {2 \sigma^2_{a,b}(t_j)} + C \equiv \\
  &\equiv -\mathcal{L}(\mathbf{x}) + C,
\end{aligned}
\label{eq:lossdata}
\end{equation}
where $V_{a,b} (t_j)$ are the model visibilities, corresponding to the image
$\mathbf{x}$, $t_j$ are the moments of time when the measurements took place,
and $C$ is some arbitrary constant that does not influence the optimization
result.

The only term left in Eq.\eqref{eq:kldiv} is $\log p(\mathbf{x})$, which
can be viewed as a problem regularizer. Despite some intuition behind it, the
choice of the exact form of the prior is, strictly speaking, arbitrary in the
general VLBI imaging setup. Also the assumption is made that the prior can be
expressed in the following way:
\begin{equation}
  \log p(\mathbf{x}) = - \sum_i \lambda_i \mathcal{R}_{i}(\mathbf{x}),
\end{equation}
where $\lambda_i$ are treated as problem hyperparameters, and the exact choice
of functions $\mathcal{R}_i$ is described in Appendix \ref{sec:apreg}. Now the
optimization problem \eqref{eq:kldiv} can be rewritten in the following form:
\begin{equation}
  \begin{aligned}
  \mathbf{\theta} \longleftarrow& \argminB_{\mathbf{\theta}}
  \mathbb{E}_{z \sim \mathcal{N}(0,\, 1)} \left[
    \mathcal{L}(\mathbf{x}) - \log \left|
        \det \frac{
          \partial \mathbf{x}
        }{\partial \mathbf{z}}
      \right| + \right. \\ &+ \left. \sum_i \lambda_i
      \mathcal{R}_{i}(\mathbf{x})
  \right] \equiv \argminB_{\mathbf{\theta}}
    \mathbb{E}_{z \sim \mathcal{N}(0,\, 1)}
    \left[\mathcal{Q}(\mathbf{z})\right].
  \end{aligned}
  \label{eq:opt}
\end{equation}

The problem Eq.\eqref{eq:opt} is a subject of stochastic gradient
optimization, since the objective function can be written in a form of
expectation \citep{Murphy2022}. So, each step is an application of an
optimization algorithm \citep[e.g AdamW, ][]{Loshchilov2017} to the gradient,
estimated on a batch of samples from the
$\mathbf{z}$ distribution:
\begin{equation}
  \nabla_{\mathbf{\theta}}
  \mathbb{E}_{z \sim \mathcal{N}(0,\, 1)} \left[\mathcal{Q}(\mathbf{z})\right]
  \underset{\mathbf{z}_k \sim \mathcal{N}(0,\, 1)}{\approx}
  \frac{1}{N} \sum_{k=1}^N \nabla_{\mathbf{\theta}}\mathcal{Q}(\mathbf{z}_k).
\end{equation}

\section{Model architecture}
\label{sec:arch}

Normalizing flows are a family of neural network architectures to parameterize
bijective continuous functions with tractable Jacobian determinant (see the
second term in Eq.\eqref{eq:opt}). They were initially proposed in
\citet{Dinh2014} along with the NICE architecture. However, in the initial
paper, mostly volume-preserving layers were suggested, which is insufficient
for an accurate density estimation \citep{Draxler2024} in Eq.\eqref{eq:dens}.
The issue can be addressed by combining Affine Coupling layers, proposed in
RealNVP \citep{Dinh2016} and invertible 1x1 convolutions, proposed in Glow
\citep{Kingma2018}.

\subsection{Conditional affine coupling layer \label{sec:coupling}}

\begin{figure}
\includegraphics[scale=0.34]{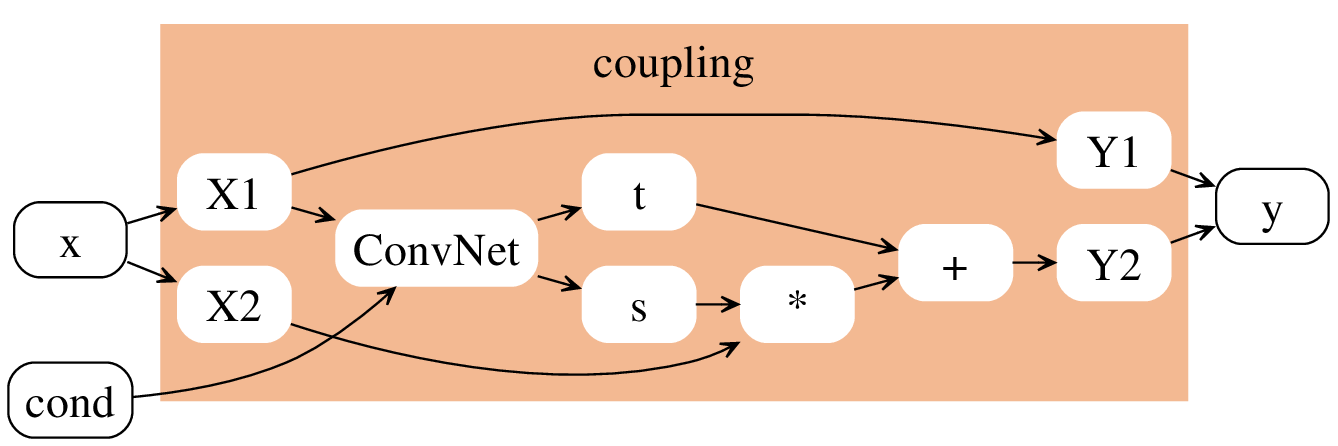}
\caption{Affine coupling block. The input is first divided into two disjoint
slices in the channel dimension or using checkerboard masks. Then, the first
slice concatenated with the conditioning are passed to a general convolutional
neural network to modify the second slice. Finally, the first slice and
modified second slice are concatenated back to produce the output.
\label{fig:archcoup}}
\end{figure}

It is a modification of the original construction for conditional inputs,
proposed in \citet{Winkler2019}. Provided a $D$ dimensional vector $\mathbf{x}$
as input, it is first split into two disjoint slices $\mathbf{x}_1$ and 
$\mathbf{x}_2$:
\begin{equation}
\begin{aligned}
  \mathbf{x}_1 &= \mathbf{x}_{1:d}, \\
  \mathbf{x}_2 &= \mathbf{x}_{d+1:D}, \\
\end{aligned}
\end{equation}
where $d < D$. Then the first input is passed to some arbitrary and possibly
non-invertible mapping $g$ as well as the conditioning $\mathbf{c}$, the result
of the mapping is split into two $D - d$ dimensional outputs $\mathbf{s}$ and
$\mathbf{t}$, used to produce the layer outputs $\mathbf{y}_1$ and
$\mathbf{y}_2$ (See Fig.~\ref{fig:archcoup}):
\begin{equation}
\begin{aligned}
  \mathbf{g} &\equiv g(\mathbf{x}_1, \mathbf{c}), \\
  \mathbf{s} &= \mathbf{g}_{1:D - d}, \\
  \mathbf{t} &= \mathbf{g}_{D - d + 1:2D - 2d}, \\
  \mathbf{y}_1 &= \mathbf{x}_1, \\
  \mathbf{y}_2 &= \mathbf{x}_2 \odot \exp(\mathbf{s}) + \mathbf{t},
\end{aligned}
\end{equation}
then both outputs are concatenated back into $D$ dimensional output
$\mathbf{y}$. The layer Jacobian matrix is block triangular with diagonal
matrices in the diagonal blocks \citep{Dinh2016}, therefore one can efficiently
compute the log-determinant:
\begin{equation}
  \log \left| \det \frac{\partial \mathbf{y}}{\mathbf{x}} \right| =
  \sum_{i=1}^{D - d} \mathbf{s}_i.
\end{equation}

Both input $\mathbf{x}$ and conditioning have a structure of an image so we
can use a convolutional neural network as the mapping $g$ thus optimizing the
model performance and leading to better overall scaling. It is denoted as
'ConvNet' in Fig.~\ref{fig:archcoup}. The specific choice of a mapping is
discussed in Appendix \ref{sec:apconv}. The input structure also influence the
choice of a way, the input is divided into $\mathbf{x}_1$ and $\mathbf{x}_2$.
Given an image with a shape of NxCxHxW, we split it in the channel dimension
wherever possible, while for single-channel inputs, we use a checkerboard mask
as proposed in \citet{Dinh2016}.

\subsection{Invertible 1x1 convolution}

Another type of invertible layers we use is invertible 1x1 convolution. It is
proposed in \citet{Kingma2018}. Assuming that the input have a shape of NxCxHxW,
it maps C channels of each pixel of the input to C channels of the
corresponding output pixel using trainable 1x1 convolutional kernel. The kernel
can be viewed as a non-degenerate CxC matrix $\mathbf{W}$ that is parameterized
using LU-decomposition:
\begin{equation}
  \mathbf{W}
  = \mathbf{P}\mathbf{L}(\mathbf{U} + \text{diag}(\exp(\mathbf{s}))),
\end{equation}
where $\mathbf{P}$ is a permutation matrix, $\mathbf{L}$ is a lower diagonal
matrix with ones on the diagonal, $\mathbf{U}$ is an upper diagonal matrix with
zeros on the diagonal and $\mathbf{L}$, $\mathbf{U}$, $\mathbf{s}$ are
trainable parameters. So, the log-determinant can be computed this way:
\begin{equation}
  \log \left| \det \frac{\partial \mathbf{y}}{\mathbf{x}} \right| =
  H W \sum_{i=1} \mathbf{s}_i.
\end{equation}

The main argument for this specific construction in the original work was
performance considerations, but these are not applicable to our case, since we
expect no more than 48 channels in the inputs (See Section~\ref{sec:wflow}).
Instead, we find it easier to initialize randomly, since any parameter value
results in invertible kernel.

\begin{figure}
\includegraphics[scale=0.25]{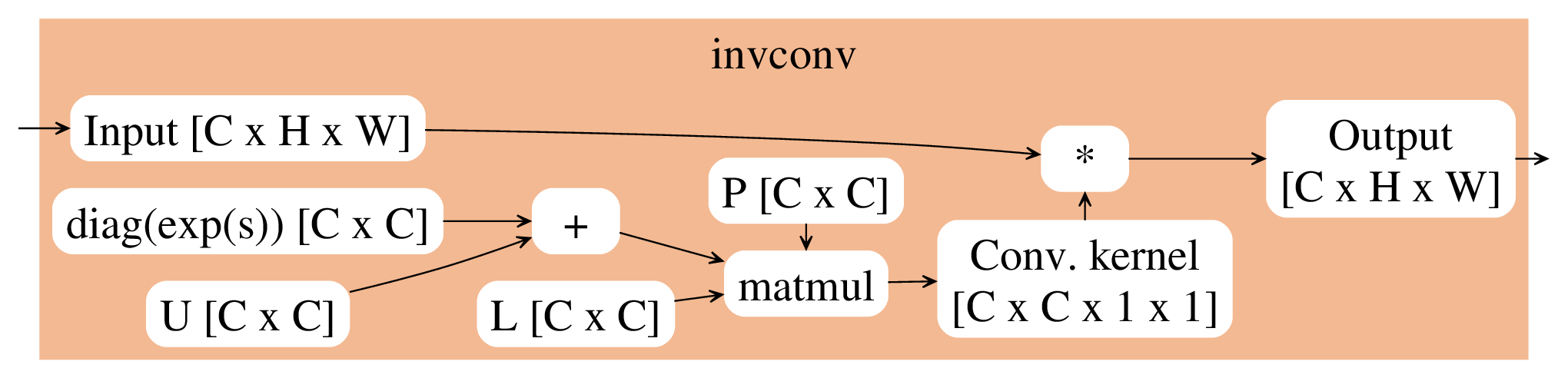}
\caption{Invertible 1x1 convolutional layer. 1x1 kernel is first assembled
from its factors and then convolved with the input image. L is a lower
triangular matrix with ones on the main diagonal, U is an upper triangular
matrix with zeros on the diagonal, s is a vector and the only parameter
influencing the total log-determinant of the transformation, P is a fixed
permutational matrix.
\label{fig:archinvconv}}
\end{figure}

\subsection{WaveletFlow}
\label{sec:wflow}

It is proven in \citet{Draxler2024}, that to get an arbitrary distribution
approximation, one can just stack a large enough number of blocks consisting of
coupling layers and trainable orthogonal transform, which can itself be
replaced with $O(1)$ linear coupling layers \citep{Koehler2021}. However, the
authors suggest that one can benefit from using more expressive normalizing
architectures in terms of convergence speed and required number of layers. In
our work we use a variation of the WaveletFlow neural network architecture
\citep{Yu2020}. The key modification here is the use of Daubechies wavelets
\citep{Daubechies1992} instead of the originally proposed Haar wavelets, since
Daubechies wavelets of higher orders (Haar wavelets are a special case with
order of 2) are known to have smaller side lobes in Fourier space. Wavelets of
orders from 6 to 16 were tested in this work. Note that we use the model in
somewhat reverse direction: we start with DWT coefficients and sample images.

\begin{figure}
\begin{minipage}[h]{0.99\linewidth}
\center{\includegraphics[scale=0.25]{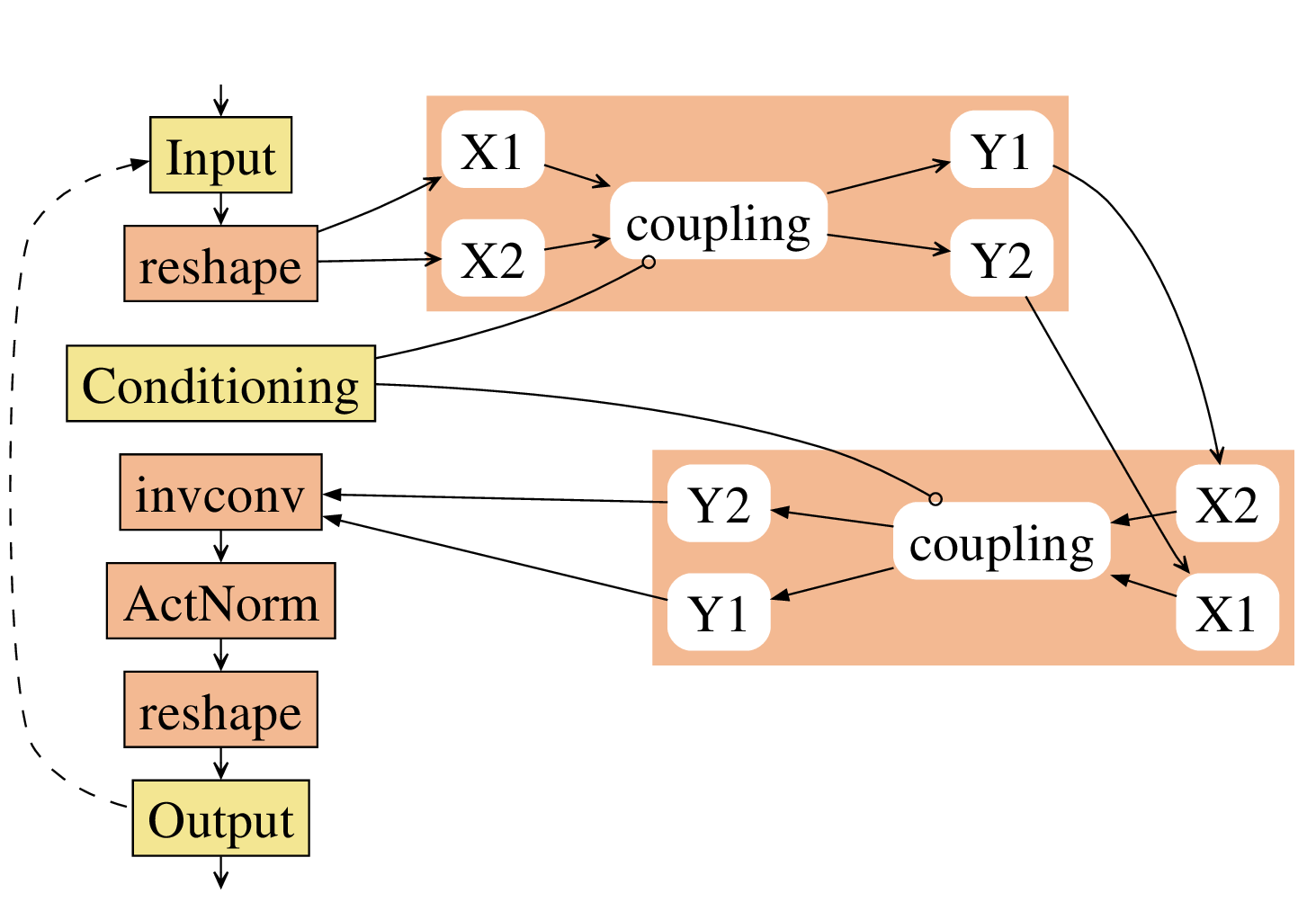}}
\end{minipage}
\caption{FlowNet architecture, which consists of repeated blocks of
invertible convolutions, couplings (optionally conditioned) and actnorm layers.
Each block starts with reshape layer: it transforms an  $N \times C \times H
\times W$ image into $N \times 16C \times H/4 \times W/4$, and ends with
another one that reshapes the image back to $N \times C \times H \times W$.
Note that in the second coupling layer, the inputs are swapped.}
\label{fig:archfull}
\end{figure}

\begin{figure}
\begin{minipage}[h]{0.85\linewidth}
\center{\includegraphics[scale=0.28]{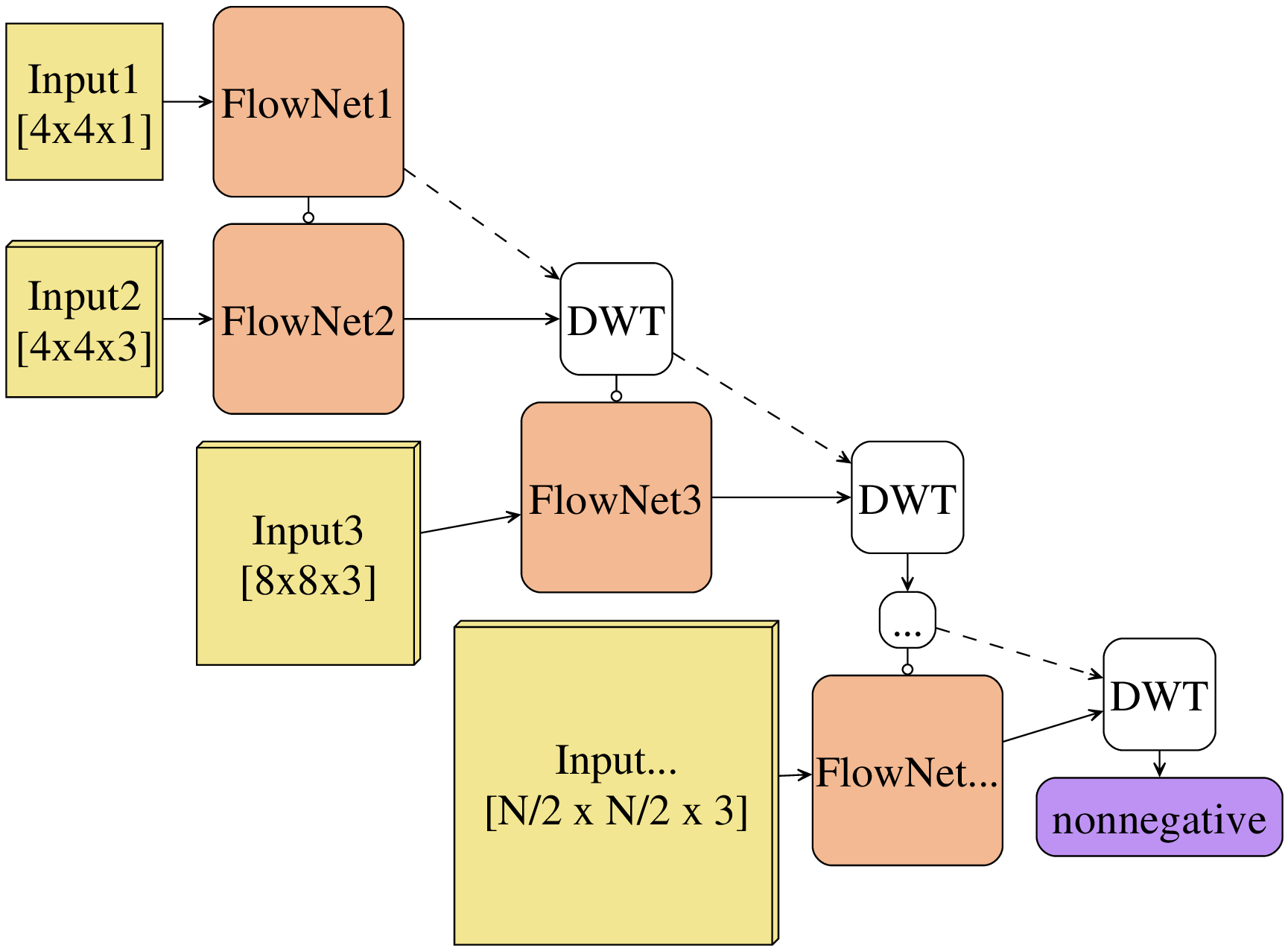}}
\end{minipage}
\caption{WaveletFlow sampling process, which starts with generating a 'coarse'
image and first DWT coefficients and results in an image that is constrained
to be positive. Arrows with circles denote conditioning.
\label{fig:archfull_sampling}}
\end{figure}

First we sample a 4x4x1 tensor from standard normal distribution and pass it to
the 'coarse' neural network (it is denoted as 'FlowNet1' in
Fig.~\ref{fig:archfull_sampling}), consisting of repeating blocks (the number
of blocks is set to 8 for all experiments) of invertible convolutions, coupling
blocks (unconditional) and actnorm layers as shown in Fig.~\ref{fig:archfull}.
Actnorm layer is a technique to stabilize training, proposed in
\citet{Kingma2018}: it acts as a trainable per-channel input rescaling
initialized so that its output is normalized right after the model weights
initialization, making it similar to batch normalization \citep{Ioffe2015}.
The result of the 'coarse' network can be viewed as a 'coarse' image (it is a
coarse version of the image in case of Haar wavelets). We then sample a 4x4x3
random tensor and pass it to the first 'coefficient' neural network
('FlowNet2' in Fig.~\ref{fig:archfull_sampling}), conditioned on the 'coarse'
image. The result is then passed to DWT along with the 'coarse' image,
producing an image with a shape of 8x8x1. The process is then repeated until
we get an image with the desired shape. Therefore, the total number of
parameters is proportional to $log_2 (H)$, where $H$ is the image size. The
full process is shown in Fig.~\ref{fig:archfull_sampling}.

Since we expect the result image to be non-negative, we need to constrain our
model. We could do this directly by adding some regularizations, but instead,
we interpret the model result as an image in logarithmic scale, which is allowed
to be negative. So the final image is an exponent of that. This slightly
differs our approach from \citet{Sun2020} where softplus function was used.
Since the logarithm of an image is used in cross-entropy regularization term
(See Appendix \ref{sec:apreg}), using raw model outputs there prevents the
loss of precision in case of negative model outputs.

\section{Experiment}

Similarly to the algorithms, radio-interferometric imaging problems can be
broadly divided into two regimes that we will refer to as 'data-intensive' and
'exploration' correspondingly. The data-intensive involves vast amounts of
visibility data to process which acts as the main constraint on the algorithms
used to solve this problem, often resulting in gridding-based algorithms. This
regime is typical for connected-element interferometry observations.
Exploration regime, on the other hand, as the name suggests, requires the
algorithm to be able to thoroughly explore the solution space and therefore
forward-modelling algorithms are often preferred since the performance impact
of consistent modelling of the visibility data is negligible. This, in turn, is
typical for (S)VLBI observations. Although our algorithm is based on forward
modelling, we test it in both regimes to demonstrate its scalability to large
data volumes without compromising its exploratory power required for SVLBI
data. In this work, we consider two interferometric array configurations, SVLBI
and connected-element interferometry, derived from two real calibrated
observational datasets.

\begin{figure}
\begin{minipage}[h]{0.48\linewidth}
\center{\includegraphics[width=1\linewidth]{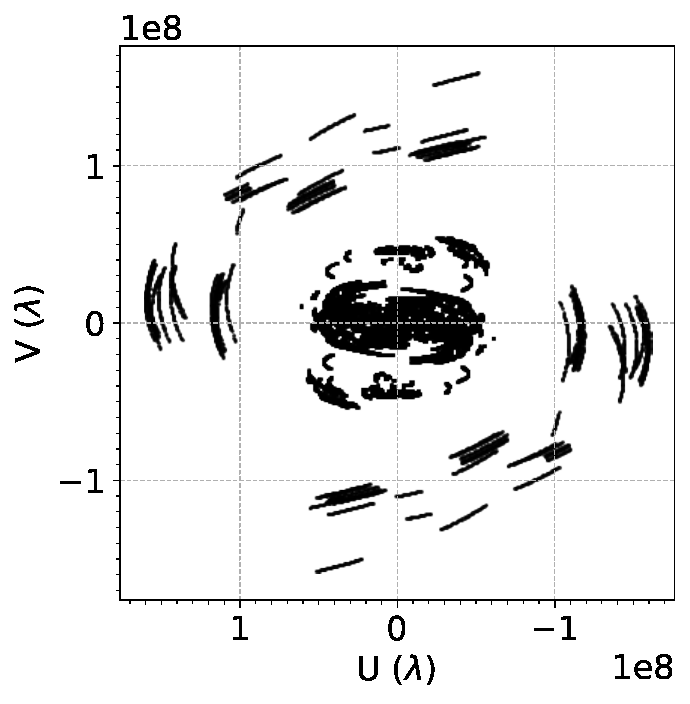}} \\
\end{minipage}
\begin{minipage}[h]{0.48\linewidth}
\center{\includegraphics[width=1\linewidth]{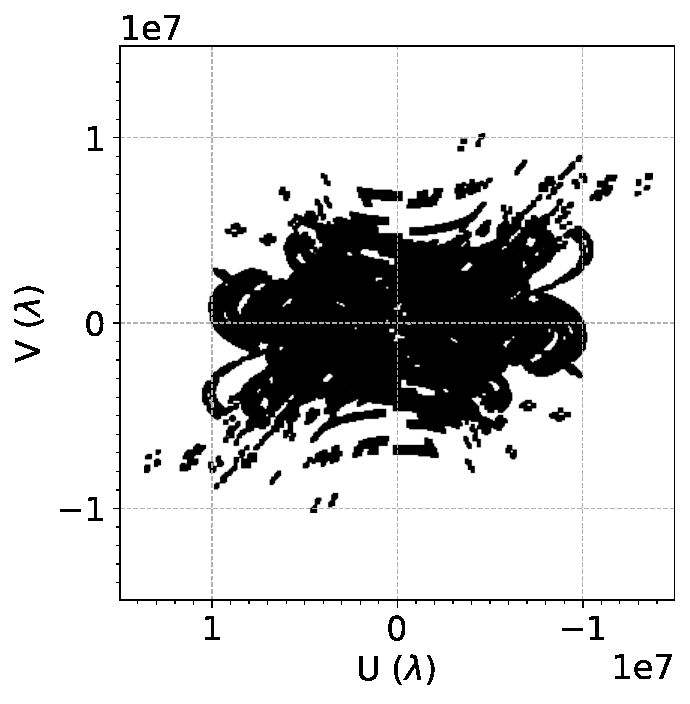}} \\
\end{minipage}
\
\caption{$(u, v)$ coverage of M87* (Left) and HL Tau (Right) is common for
both real and model observations, that is, the coordinates of the  Fourier
components, measured during the observations. Spatial frequencies are measured
in units of wavelength.
}
\label{fig:uvcoverage}
\end{figure}

SVLBI observations were conducted on RadioAstron and 27 ground-based stations
on June 4, 2014 with the total experiment duration of about 25 hours. The total
bandwidth is 32 MHz with the central frequency of 1652.25 MHz (L band), only LL
polarization is selected. The achieved UV-coverage is shown in
Fig.~\ref{fig:uvcoverage}.

One may notice in \eqref{eq:kldiv}, \eqref{eq:lossdata}, that the method
requires careful visibilities error estimation even for the 'convex' case to
assign proper weight to the log-determinant term of \eqref{eq:kldiv}. The
standard deviation of the thermal noise components is estimated in the
following way:
\begin{equation}
\sigma_{a,b} = \frac{1}{\eta} \sqrt{
  \frac{SEFD_a SEFD_b}{2 \Delta \nu \tau}
},
\label{eq:sigma_real}
\end{equation}
where $\eta$ denotes the quantization factor (it is 0.88 in the case of the
2-bit quantization), $\Delta \nu$ is the width of the frequency band, $\tau$ is
the integration time and $SEFD_{a}$, $SEFD_{b}$ are the system equivalent flux
densities for the stations $a$ and $b$, respectively, at given frequency,
polarization and direction. Note that the integration time $\tau$, is usually
proportional to the weights of the visibilities on the correlator output. In
turn, $SEFD_a$ for each station $a$ is a function of the angular coordinates
of the object, frequency, time and selected polarization and usually reported
in the AIPS ANTAB format \citep{Wells1985}:
\begin{equation}
SEFD_a(\nu, t, m) =
\frac{T_{sys}(\nu, t, m)}{DPFU(m) \cdot P(A(t), h(t))},
\end{equation}
where $m$ is the selected polarization, $t$ is time, and $A(t)$, $h(t)$, $P$
are angular coordinates of the object and some polynomial of coordinates.

The SVLBI dataset is used in two settings: channel- and time- averaged, and
unaveraged data, consisting of $\approx 4 \times 10^8$ points. While both are
suitable for imaging the calibrated data (provided that the gain calibration
is precise) and both require thorough exploration of the solution space, they
enable us to test how different parts of the algorithm will scale separately.
The former is small enough to be modelled entirely during the algorithm
iteration, and is used to test the algorithm precision (see
Section~\ref{sec:modeldata}) and the neural architecture scaling properties
(see Section~\ref{sec:pixelscaling}). On the other hand, the latter is an
important case of data-intensive regime which is mainly explored in
Section~\ref{sec:datascaling}. Although, as was stated above, this regime is
typical for connected-interferometry setups, algorithms used in
self-calibration are usually also constrained by the noise-levels and size of
the dataset before the averaging. Error estimates (Eq.\eqref{eq:sigma_real})
are adjusted to correspond to the effective channel widths and integration
times.

\begin{figure}
\begin{minipage}[h]{0.43\linewidth}
\center{\includegraphics[width=1\linewidth]{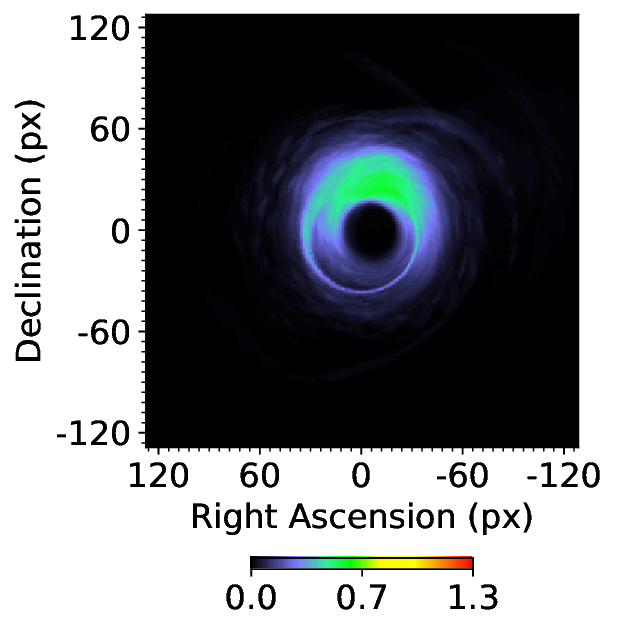}} \\
\end{minipage}
\hfill
\begin{minipage}[h]{0.43\linewidth}
\center{\includegraphics[width=1\linewidth]{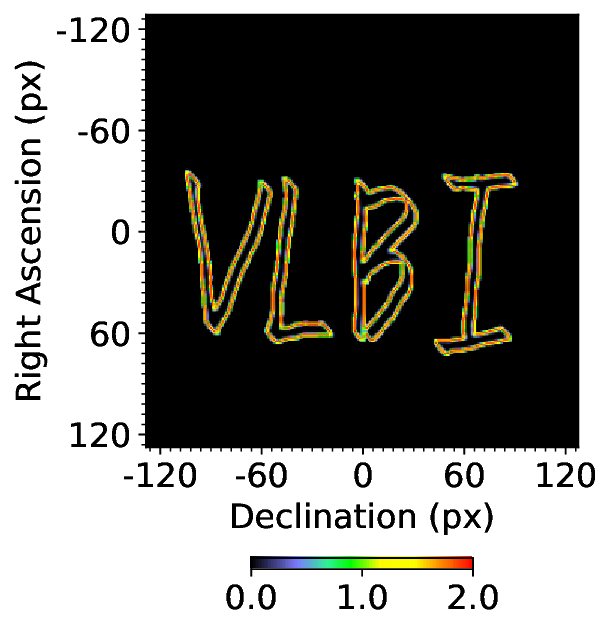}} \\
\end{minipage}
\caption{Model sources: black hole accretion disk image on the left, 'VLBI'
label on the right. All fluxes are in the units of millijanskys.}
\label{fig:modelmaps}
\end{figure}

The connected-element interferometry is represented by pipeline-calibrated ALMA
Science Verification data of HL Tau observations in Band 7 \citep{Alma2015}
consisting of $\approx 4 \times 10^7$ valid visibility data points across all
frequency channels with total bandwidth of 8 GHz centred at 344 GHz.
Polarization I is used. The UV-coverage is shown in Fig.~\ref{fig:uvcoverage}.
Thermal error standard deviations are provided within the dataset. While being
relatively small compared to the largest datasets produced by the largest
arrays, it is an important case of the data-intensive regime, when the whole
dataset still can fit into the computer RAM and wide-field imaging techniques
are not required.

Despite using synthetic datasets in the tests discussed in the next
sections, it is important to demonstrate that the algorithm is capable of
imaging the real observations, where the precise error distribution is unknown
and only some parameter estimates are available, and calibration is not exact.
The result logscale image of 256x256 pixels and the central region of 2048x2048
map corresponding to RadioAstron and ALMA datasets respectively are shown in
Fig.~\ref{fig:realdata} along with baseline images. CLEAN result is used
as the baseline for M87 data, while the baseline map for HL Tau is provided
by the ALMA dataset itself.

\begin{figure*}
\begin{minipage}[c]{0.003\linewidth}
  \centering\rotatebox{90}{\textbf{\small avg. NF}}
\end{minipage}\hfill
\begin{minipage}[h]{0.18\linewidth}
\center{\includegraphics[width=1\linewidth]{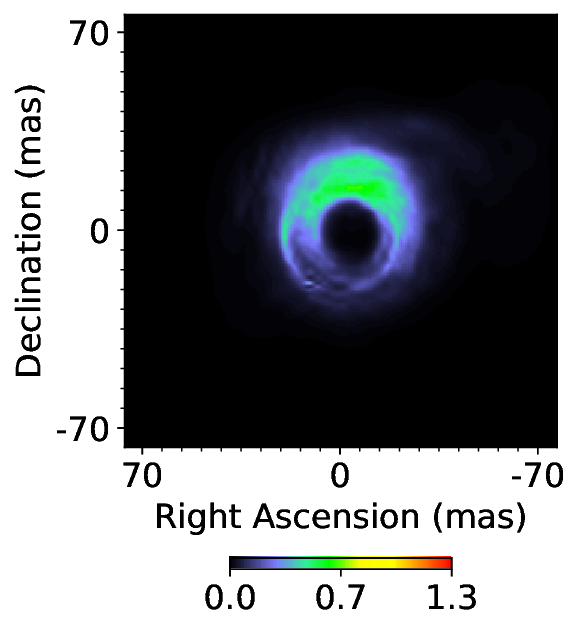}} \\
\end{minipage}
\hfill
\begin{minipage}[h]{0.18\linewidth}
\center{\includegraphics[width=1\linewidth]{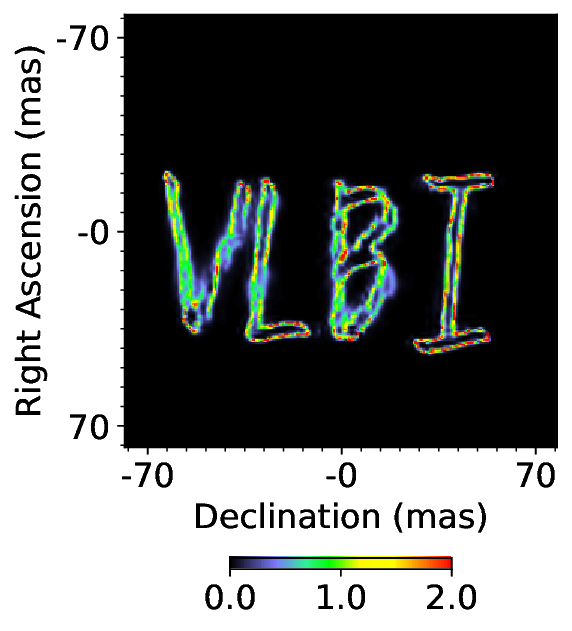}} \\
\end{minipage}
\hfill
\begin{minipage}[h]{0.18\linewidth}
\center{\includegraphics[width=1\linewidth]{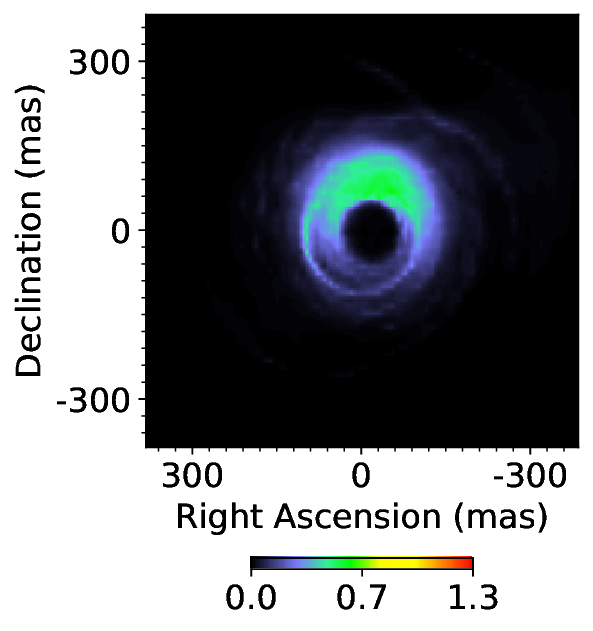}} \\
\end{minipage}
\hfill
\begin{minipage}[h]{0.18\linewidth}
\center{\includegraphics[width=1\linewidth]{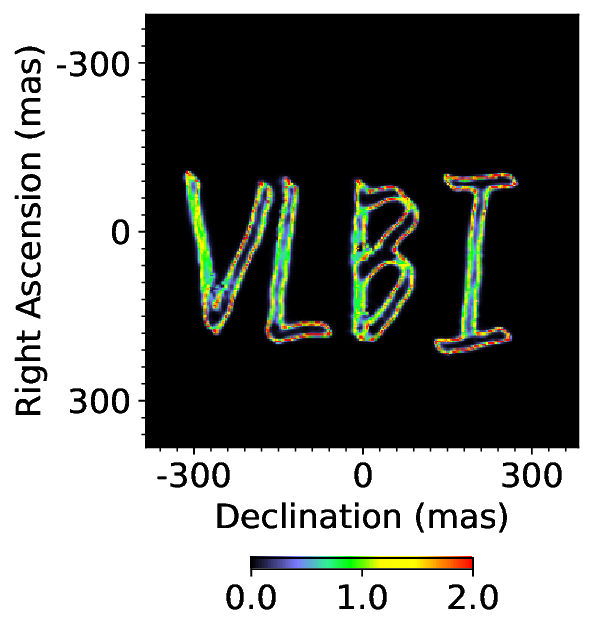}} \\
\end{minipage}
\vfill
\begin{minipage}[c]{0.003\linewidth}
  \centering\rotatebox{90}{\textbf{\small CLEAN}}
\end{minipage}\hfill
\begin{minipage}[h]{0.18\linewidth}
\center{\includegraphics[width=1\linewidth]{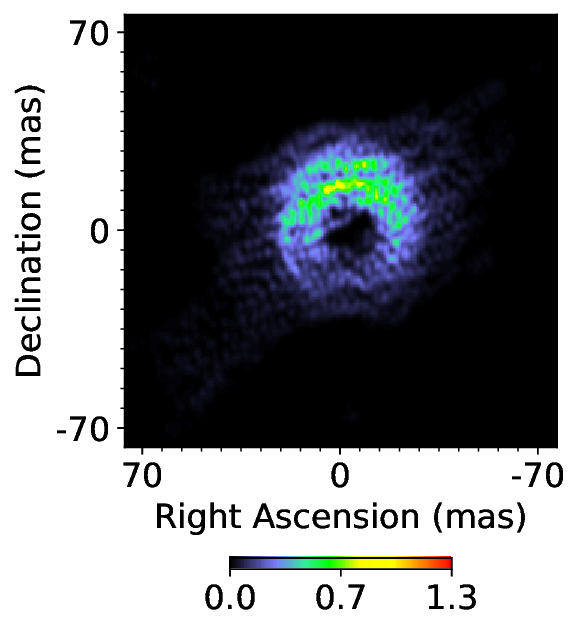}} \\
\end{minipage}
\hfill
\begin{minipage}[h]{0.18\linewidth}
\center{\includegraphics[width=1\linewidth]{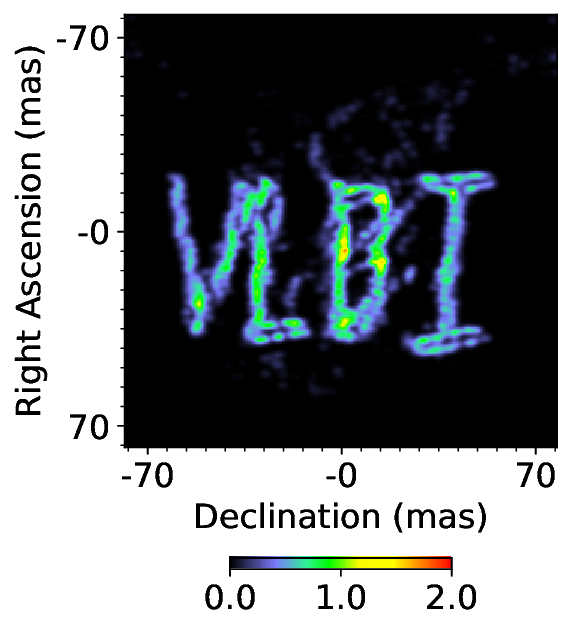}} \\
\end{minipage}
\hfill
\begin{minipage}[h]{0.18\linewidth}
\center{\includegraphics[width=1\linewidth]{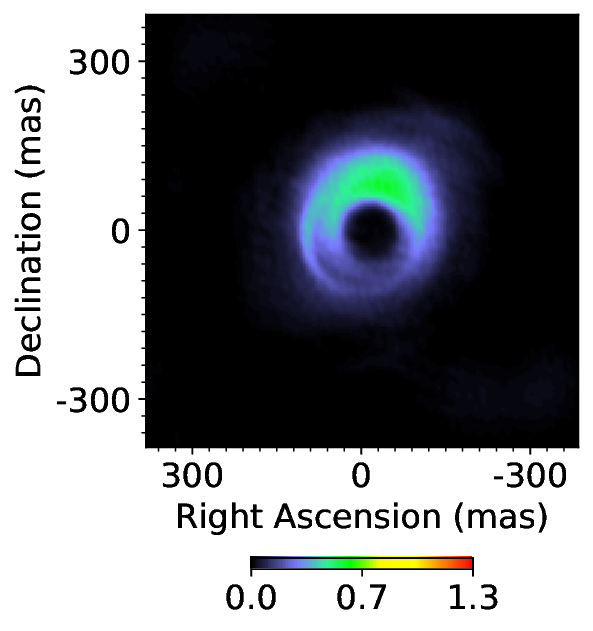}} \\
\end{minipage}
\hfill
\begin{minipage}[h]{0.18\linewidth}
\center{\includegraphics[width=1\linewidth]{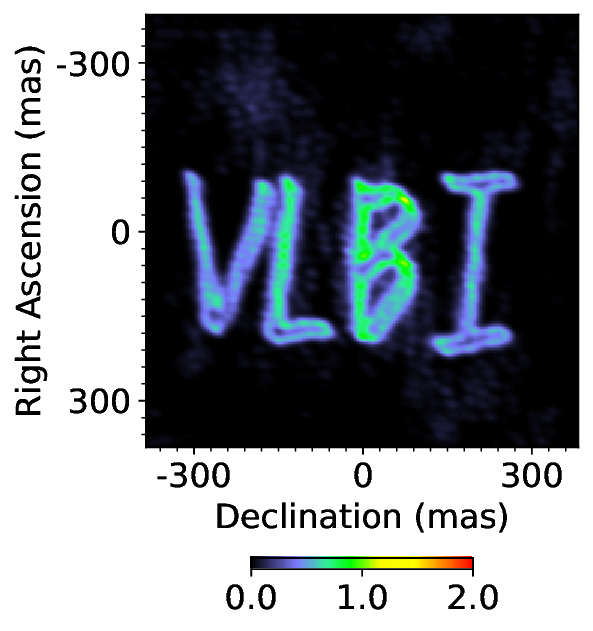}} \\
\end{minipage}
\vfill
\begin{minipage}[c]{0.003\linewidth}
  \centering\rotatebox{90}{\textbf{\small NF residual}}
\end{minipage}\hfill
\begin{minipage}[h]{0.18\linewidth}
\center{\includegraphics[width=1\linewidth]{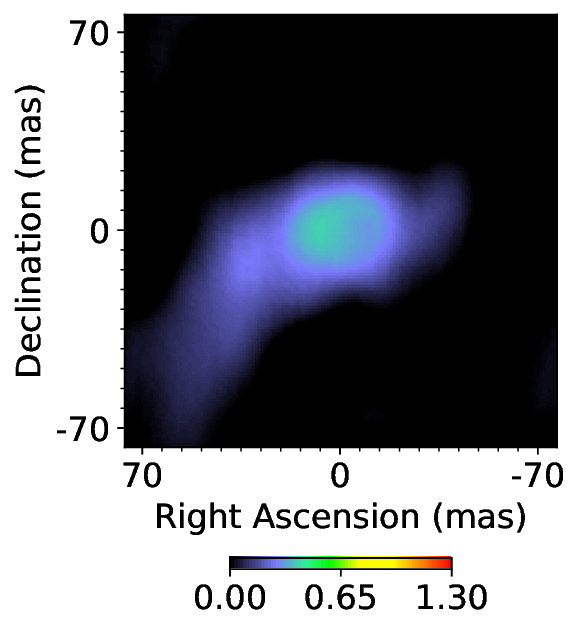}} \\
\end{minipage}
\hfill
\begin{minipage}[h]{0.18\linewidth}
\center{\includegraphics[width=1\linewidth]{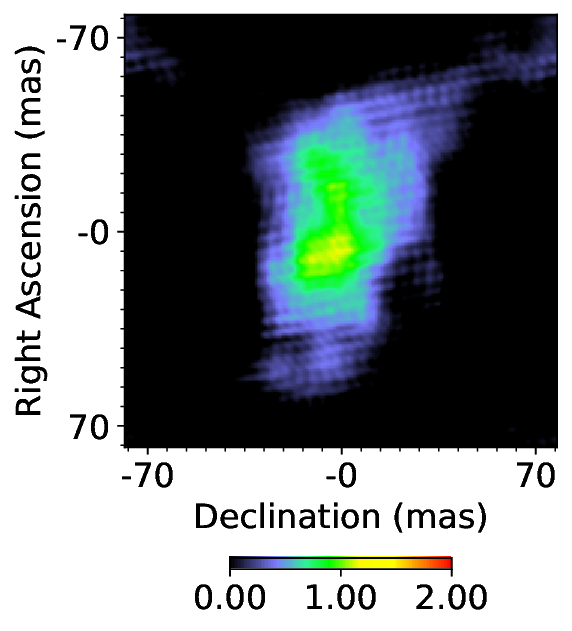}} \\
\end{minipage}
\hfill
\begin{minipage}[h]{0.18\linewidth}
\center{\includegraphics[width=1\linewidth]{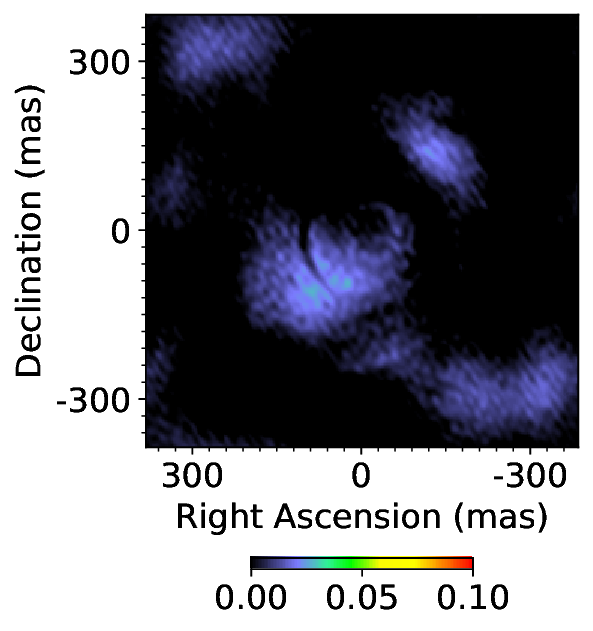}} \\
\end{minipage}
\hfill
\begin{minipage}[h]{0.18\linewidth}
\center{\includegraphics[width=1\linewidth]{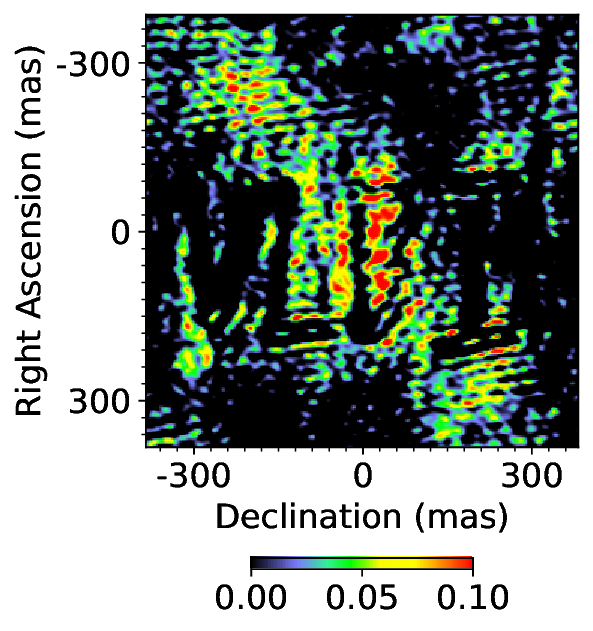}} \\
\end{minipage}
\vfill
\begin{minipage}[c]{0.003\linewidth}
  \centering\rotatebox{90}{\textbf{\small Model residual}}
\end{minipage}\hfill
\begin{minipage}[h]{0.18\linewidth}
\center{\includegraphics[width=1\linewidth]{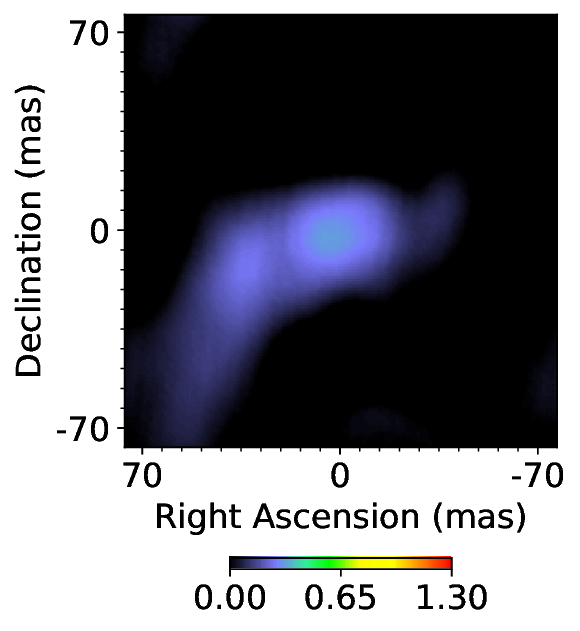}} \\
\end{minipage}
\hfill
\begin{minipage}[h]{0.18\linewidth}
\center{\includegraphics[width=1\linewidth]{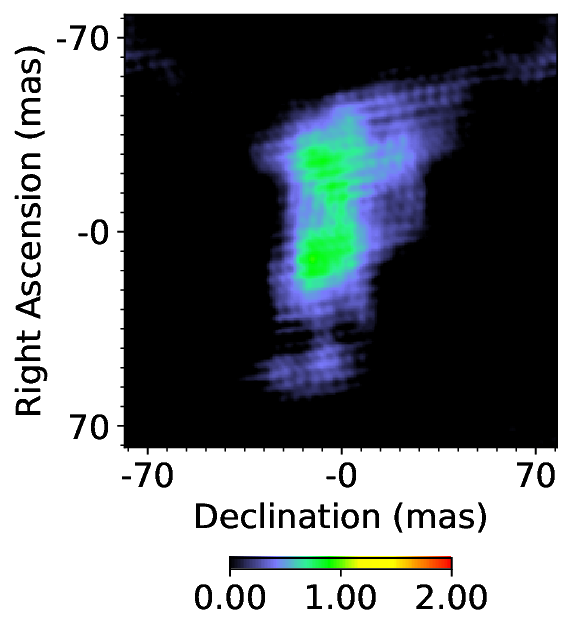}} \\
\end{minipage}
\hfill
\begin{minipage}[h]{0.18\linewidth}
\center{\includegraphics[width=1\linewidth]{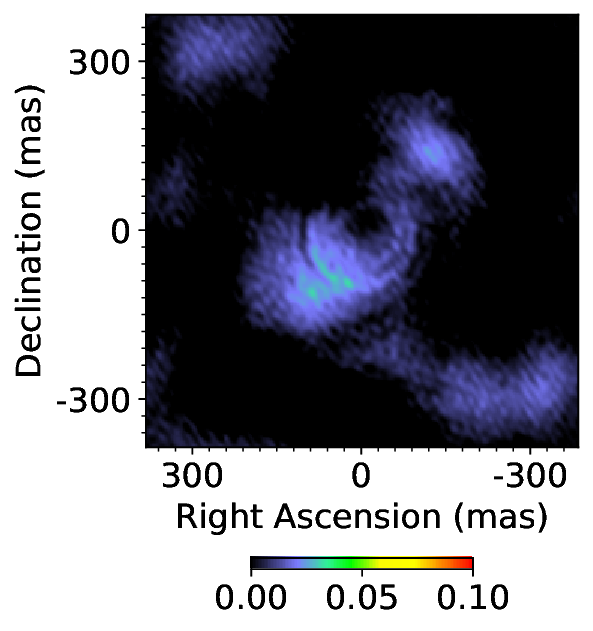}} \\
\end{minipage}
\hfill
\begin{minipage}[h]{0.18\linewidth}
\center{\includegraphics[width=1\linewidth]{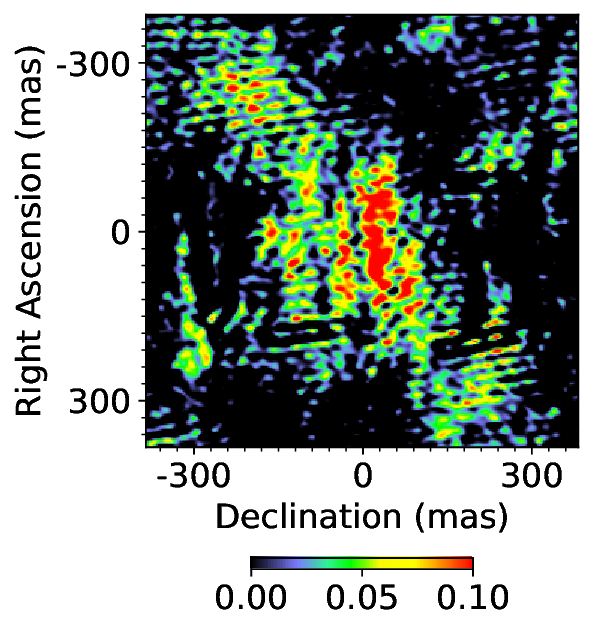}} \\
\end{minipage}
\vfill
\begin{minipage}[c]{0.003\linewidth}
  \centering\rotatebox{90}{\textbf{\small NF FFT std}}
\end{minipage}\hfill
\begin{minipage}[h]{0.18\linewidth}
\center{\includegraphics[width=1\linewidth]{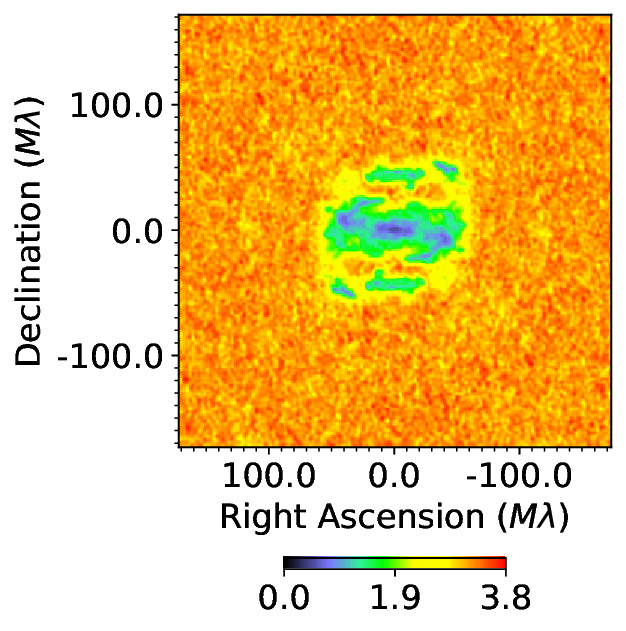}} \\
\end{minipage}
\hfill
\begin{minipage}[h]{0.18\linewidth}
\center{\includegraphics[width=1\linewidth]{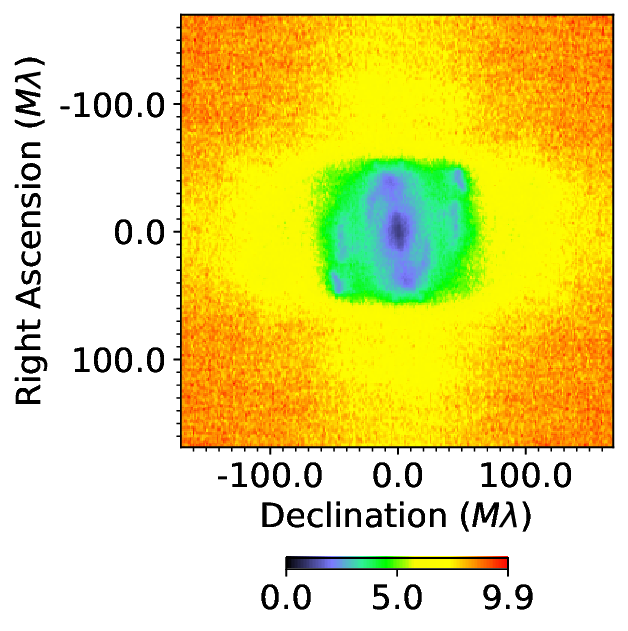}} \\
\end{minipage}
\hfill
\begin{minipage}[h]{0.18\linewidth}
\center{\includegraphics[width=1\linewidth]{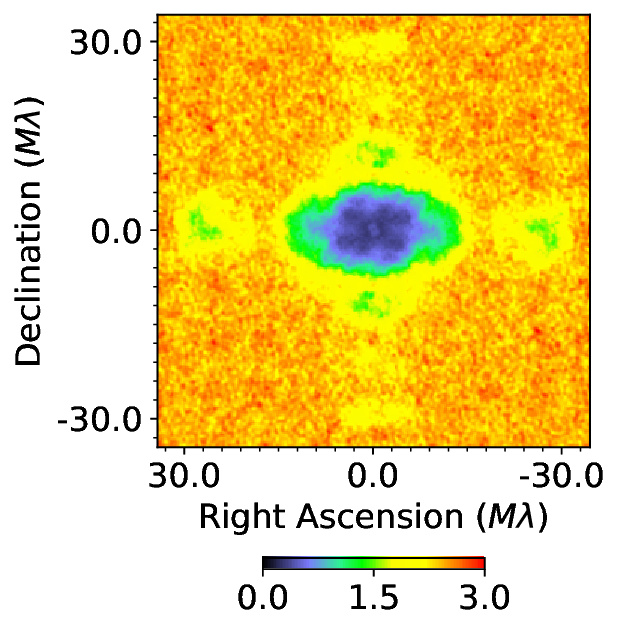}} \\
\end{minipage}
\hfill
\begin{minipage}[h]{0.18\linewidth}
\center{\includegraphics[width=1\linewidth]{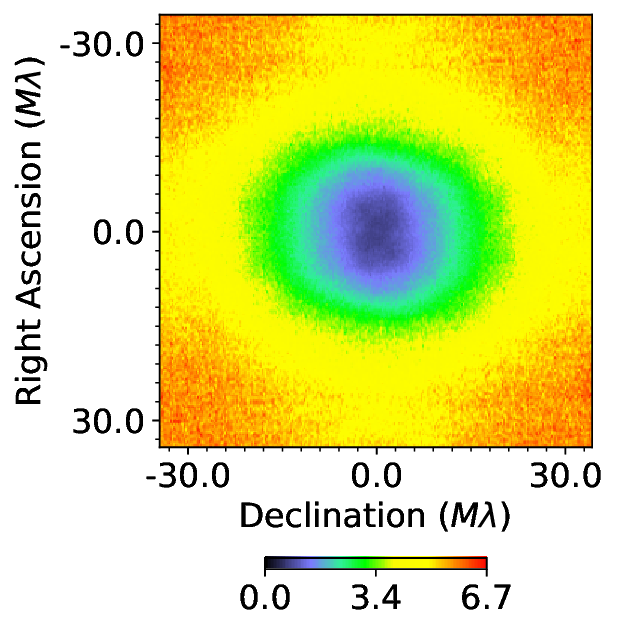}} \\
\end{minipage}
\vfill
\begin{minipage}[c]{0.003\linewidth}
  \centering\rotatebox{90}{\textbf{\small MAP}}
\end{minipage}\hfill
\begin{minipage}[h]{0.18\linewidth}
\center{\includegraphics[width=1\linewidth]{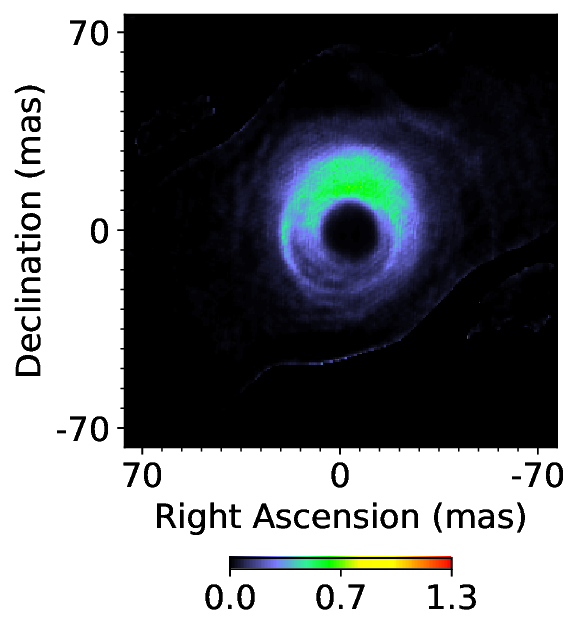}} \\
\end{minipage}
\hfill
\begin{minipage}[h]{0.18\linewidth}
\center{\includegraphics[width=1\linewidth]{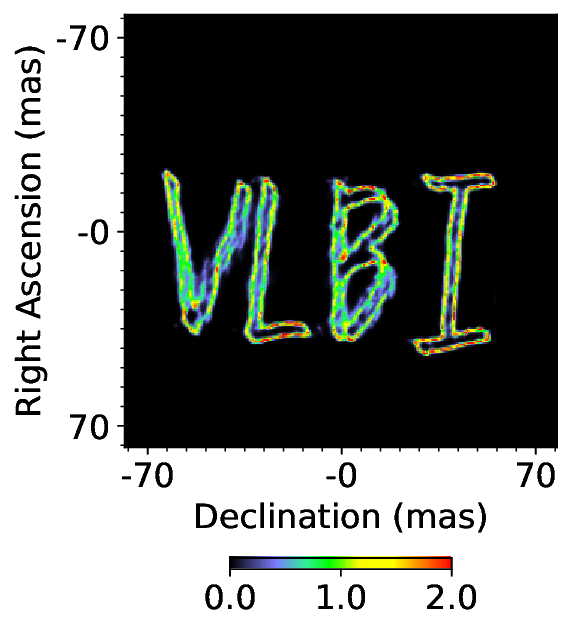}} \\
\end{minipage}
\hfill
\begin{minipage}[h]{0.18\linewidth}
~
\end{minipage}
\hfill
\begin{minipage}[h]{0.18\linewidth}
~
\end{minipage}
\vfill

\caption{Model data imaging. The first two columns and the last two columns
correspond to the M87* and HL Tau $(u, v)$ coverage respectively.
(First row) Normalizing flow results, averaged
over the generated sample of 32 images.
(Second row) CLEAN algorithm results, that is, CLEAN maps.
(Third row) The residual corresponding to the averaged NF result.
(Fourth row) The residual corresponding to the model map itself.
(Fifth row) The standard deviation of the Fourier coefficients across the sample
of the NF results.
(Sixth row) MAP algorithm simulation result.
All fluxes are in the units of millijanskys.}
\label{fig:modeldata}
\end{figure*}

\subsection{Model data}
\label{sec:modeldataalgo}

\begin{figure}
\begin{minipage}[h]{0.43\linewidth}
\center{\includegraphics[width=1\linewidth]{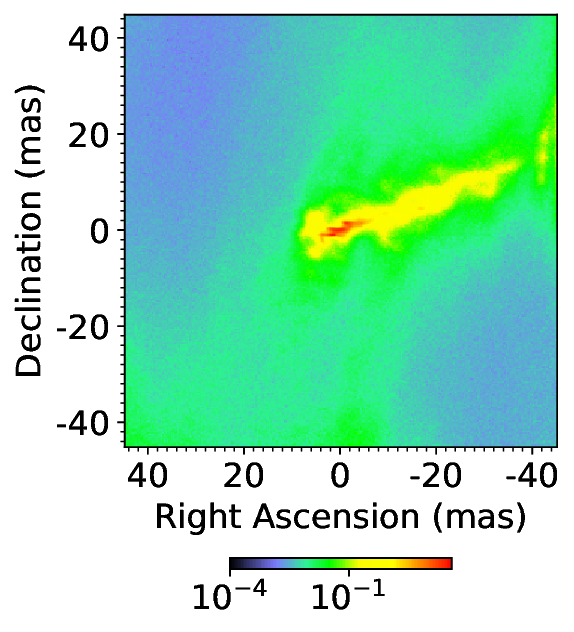}} \\
\end{minipage}
\hfill
\begin{minipage}[h]{0.43\linewidth}
\center{\includegraphics[width=1\linewidth]{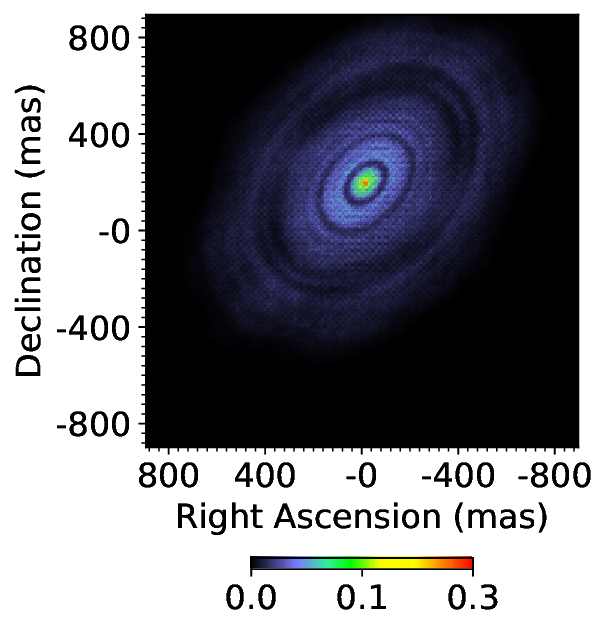}} \\
\end{minipage}
\vfill
\begin{minipage}[h]{0.43\linewidth}
\center{\includegraphics[width=1\linewidth]{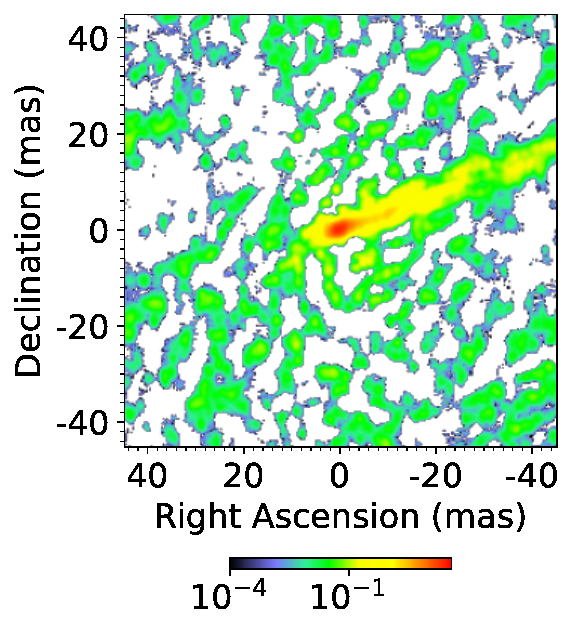}} \\
\end{minipage}
\hfill
\begin{minipage}[h]{0.43\linewidth}
\center{\includegraphics[width=1\linewidth]{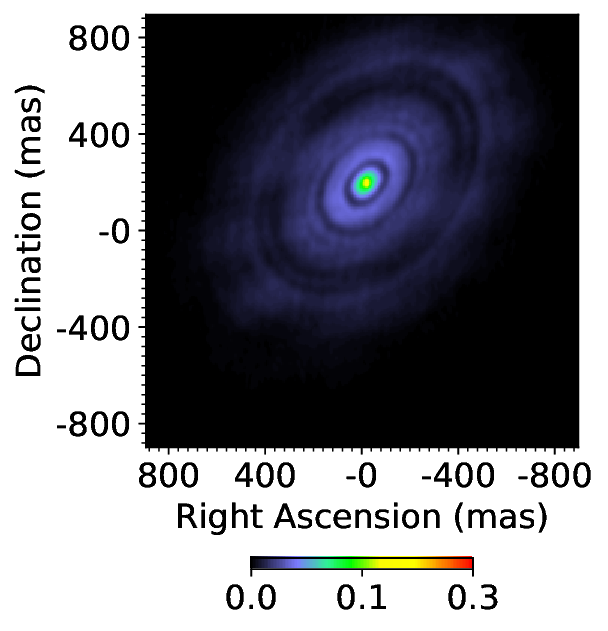}} \\
\end{minipage}
\caption{
Real calibrated observational data imaging.
(Top row) Sample average images from the algorithm. One may notice the
artefacts at the M87* image boundaries which are probably caused
by the DWT assumption that the signal is periodic.
(Bottom row) Baseline images: CLEAN result of the M87* dataset and
the image map of HL Tau provided within the dataset.
(Left column) Images of M87* (log scale).
(Right column) Images of HL Tau, that is, the centre of 2048x2048 map.}
\label{fig:realdata}
\end{figure}

SVLBI techniques usually impose some limitations on the candidate objects to be
observed. It is common to assume that the sources of most interest are AGNs
that are known to have some compact and bright structures. For example, in
SVLBI dataset, RadioAstron project M87* jet observational data are used
(see Fig.~\ref{fig:realdata}). However, we test the algorithm on both
astrophysical and non-astrophysical image maps to ensure that the prior
assumptions we use don't imply jet-like features.

The model images are shown in Fig.~\ref{fig:modelmaps}. It is important to
note, that the simulated accretion disk image, although it is astrophysical,
cannot be resolved with UV-coverages of the considered datasets, as it would
require microarcsecond-scale resolution. However, it still represents
a target of VLBI observations. Instead, for all experiments, the model sources
are scaled in a way that the angular size of the pixel remains the same as on
the reference images in Fig.~\ref{fig:realdata}, i.e. 256 pixels on the images
correspond to $153$ mas and $768$ mas for RadioAstron and HL Tau datasets
respectively.

Model observations use the same $(u, v)$ coordinates as the real data, but the
values of the visibilities are altered to correspond to the model images:
\begin{equation}
  V_{alter}(u_i, v_i) = V(u_i, v_i) + \epsilon_{i},
\end{equation}
where $\epsilon_i$ is independent symmetric Gaussian complex noise with
standard deviations taken from the estimates \eqref{eq:sigma_real}, and
$V(u, v)$ are 'true' visibilities:
\begin{equation}
  V(u, v) = \iint I(x, y) e^{2 \pi i (u x + v y)} dx dy.
\end{equation}
In this work, image spectral curvature is not modelled, however, it is
important to note, that different frequency channels produce slightly different
sets of $(u, v)$ coordinates, i.e. $u,v \propto f$, where $f$ is a frequency,
and this phenomenon is taken into account.

Since the model image consists of discrete pixels, one needs to create its
continuous representation. We model this representation as a convolution of the
discrete image with the pulse function $p(x, y)$:
\begin{equation}
\begin{aligned}
  I(x, y) &= p(x, y) \ast \sum_{m, n} I_{m, n}
  \delta(x - m \Delta x, y - n \Delta y), \\
  V(u, v) &= \tilde{p}(u \Delta x, v \Delta y) \sum_{m,n} I_{m, n}
  e^{2 \pi i (u m \Delta x + v n \Delta y)},
\end{aligned}
\end{equation}
In this work, rectangular pulses are used:
\begin{equation}
  \tilde{p}(u, v) = \text{sinc}(u) \text{sinc}(v).
\end{equation}

\subsection{Model data imaging fidelity}
\label{sec:modeldata}

Then, we evaluate the algorithm to obtain the samples of images for both array
configuration (averaged M87* and HL Tau) and both models, making four test
cases in total. For each test case, we take one image from a sample at random
and also calculate an averaged image over the sample to compare both to the
model source map. For a baseline, we use CLEAN algorithm from ASL software
package \citep{Likhachev2020} in a setup, where all components are restricted
to be non-negative. For the RadioAstron data, natural weighting is used as the
CLEAN algorithm approximates the maximum likelihood estimate. In contrast, the
uniform weighting is applied to HL Tau observations since it both better aligns
with the original data calibration script and the visibility data are
concentrated towards the centre. For both regimes, CLEAN components convolved
with their corresponding clean beams are used for the comparison. Although the
CLEAN components themselves are the solution to the imaging problem, they are
too sparse for the metrics used in this work. Finally, for RadioAstron data a
MAP-like algorithm is simulated via optimizing the objective function
(Eq.\eqref{eq:opt}) with respect to the image map with the log-determinant
term omitted.

The comparison is conducted using MSE metric and SSIM index
\citep{Wang2004,deepmind2020jax}:
\begin{equation}
\begin{aligned}
MSE(\mathbf{x}, \mathbf{y}) &= \frac{\sum_{i,j} (x_{i,j} - y_{i,j})^2}{H W}, \\
SSIM(\mathbf{x}, \mathbf{y}) &=
\frac{(2\mu_x\mu_y + c_1)(2\sigma_{xy} + c_2)}
{(\mu_x^2 + \mu_y^2 + c_1)(\sigma_x^2 + \sigma_y^2 + c_2)} \\
c_1 &= (k_1 L)^2,\: c_2 = (k_2 L)^2,\: k_1 = 0.01,\: k_2 = 0.03
\end{aligned}
\label{eq:ssimmse}
\end{equation}
where $H, W$ denote image sizes, $L$ is the dynamic range, $\mu_x$, $\mu_y$
and $\sigma_x^2$, $\sigma_y^2$ are averages and variances across the
$\mathbf{x}$ and $\mathbf{y}$ and $\sigma_{xy}$ is the covariance of the images.
The results are summarized in
Table~\ref{tab:metrics} and the result images are shown in
Fig.~\ref{fig:modeldata}. Both NF and MAP perform similarly in terms of the
metrics, however the sample average image performs
consistently better for both models and both outperform the CLEAN results.
In terms of SSIM index, even individual NF image realization performs slightly
better than the CLEAN results.

\begin{table}%
\caption{Reconstruction fidelity comparison between NF, MAP and CLEAN
('non-negative' version of the algorithm).
\label{tab:metrics}}
\begin{tabular}{lllllll}
 & Model & Metric & NF & NF avg. & CLEAN & Map \\
\hline
M87* & Disk & MSE & 2.64e-10 & 1.11e-10 & 8.49e-10 & 8.94e-11 \\
 & Disk & SSIM & 0.860 & 0.899 & 0.673 & 0.918 \\
 & VLBI & MSE & 3.34e-08 & 3.13e-08 & 7.20e-08 & 3.04e-08 \\
 & VLBI & SSIM & 0.894 & 0.899 & 0.705 & 0.913 \\
\hline
HL Tau & Disk & MSE & 1.53e-10 & 6.16e-11 & 2.08e-10 & -- \\
 & Disk & SSIM & 0.913 & 0.948 & 0.696 & -- \\
 & VLBI & MSE & 2.59e-08 & 2.56e-08 & 6.72e-08 & -- \\
 & VLBI & SSIM & 0.947 & 0.948 & 0.544 & -- \\
\end{tabular}
\end{table}

\subsection{Exploration regime scaling}
\label{sec:pixelscaling}

Then, we explore the scaling properties of the algorithms as the number of pixels
in recovered images grows. We use 2 hardware configurations: NVIDIA RTX 4070 Ti
GPU and a cluster node with 4 interconnected NVIDIA Tesla A100 GPUs
(distributed data parallel training scheme is used for simplicity and
scalability for relatively small neural networks). The averaged M87* array
configuration is used for all tests.

The WaveletFlow architecture is implemented using JAX \citep{Frostig2019}
framework. Each block (see Section~\ref{sec:wflow}) uses gradient checkpointing
technique \citep{Chen2016} to lower the memory usage. To both lower memory usage
and required memory throughput, 16-bit floats are used to evaluate the ConvNet
subnets (see Section~\ref{sec:coupling} and Appendix~\ref{sec:apconv}).
The sample is spread evenly across GPUs. The maximum number of model parameters
was $\approx 10^8$.

For comparison, Glow and RealNVP architectures, based on the implementation
by \citet{Sun2020}, were reimplemented using the JAX framework. To reduce the
memory footprint, the number of channels in the convolutional kernels of the
Glow blocks was decreased from 512 to 256. The number of blocks is set to 4
by default, although 8-block setup is tested for larger maps on the cluster
node.

In contrast with Section~\ref{sec:modeldata}, the number of iterations for each
algorithm was limited to 20000. Results of each run are summarized in
Table~\ref{tab:scaling} along with the achieved imaging fidelity metrics
(Eq. \eqref{eq:ssimmse}) values. For each image size above 256x256, the fixed
angular resolution is maintained and only the centre is compared to the ground
truth. In contrast, the images with sizes below 256x256 are compared to the
averaged model map.

Notably, RealNVP is slightly faster for $64 \times 64$ maps, however, this
result does not scale because the architecture employs fully-connected layers,
which becomes computationally prohibitive for image sizes exceeding
$\approx 100 \times 100$ pixels. In all tests, WaveletFlow algorithm
outperformed Glow architecture both in terms of image fidelity and the time,
provided the computational budget. The Glow architecture implementation does
not scale beyond $512 \times 512$ maps on the PC and beyond $1024 \times 1024$
for 8-block setup for the cluster node because of the memory footprint,
although this can be potentially improved using different types of distributed
parallelism techniques instead of DDP (usually compromising the speed), or
with ConvNet subnets adjustments. The cluster hardware, being faster than the
PC in terms of total GPU memory bandwidth and compute (the same experiment with
$512 \times 512$ WaveletFlow maps is conducted twice for the comparison),
enable us to test even larger models and maps. Interestingly, the 4-block Glow
setup was insufficient for $512 \times 512$ maps even for larger sample sizes
(combined with and increased learning rate, this could lead to faster
convergence): the 8-block architecture, which is closer to the WaveletFlow with
one block per scale, yields better fidelity metric results. While the algorithm
performs worse than the MAP-based approach in this experiment, it remains a full
sampling method capable of imaging relatively large maps on a PC.

\begin{table*}%
\caption{The scalability of the algorithms. \label{tab:scaling}}
\begin{tabular}{rllrlrrrrr}
npix & hardware & algorithm & sample & duration & SSIM (VLBI) & MSE (VLBI) & iter/s  & SSIM (Disk) & MSE (Disk)\\
\hline
64 & 1 x RTX 4070 Ti & RealNVP & 32 & 0:14:33 & 0.69 & 8.00e-06 & 22.88 & 0.94 & 6.02e-08 \\
64 & 1 x RTX 4070 Ti & WaveletFlow & 32 & 0:22:26 & 0.70 & 7.50e-06 & 14.84 & 0.92 & 1.70e-07\\
128 & 1 x RTX 4070 Ti & Glow & 8 & 1:06:36 & 0.88 & 6.19e-07 & 5.00 & 0.91 & 2.66e-09\\
128 & 1 x RTX 4070 Ti & WaveletFlow & 32 & 0:33:17 & 0.88 & 5.51e-07 & 10.00& 0.91 & 1.30e-08\\
256 & 1 x RTX 4070 Ti & Glow & 8 & 3:45:53 & 0.87 & 6.25e-08 & 1.47 & 0.79 & 1.16e-09\\
256 & 1 x RTX 4070 Ti & MAP & 32 & 0:06:23 & 0.91 & 2.97e-08 & 51.68 & 0.92 & 8.94e-11\\
256 & 1 x RTX 4070 Ti & WaveletFlow & 32 & 0:53:14 & 0.89 & 3.38e-08 & 6.25 & 0.87 & 2.49e-10\\
512 & 1 x RTX 4070 Ti & Glow & 2 & 3:41:58 & 0.88 & 5.90e-08 & 1.50 & 0.74 & 2.77e-09\\
512 & 1 x RTX 4070 Ti & WaveletFlow & 32 & 2:25:12 & 0.89 & 4.05e-08 & 2.29& 0.83 & 8.84e-10\\
512 & 4 x A100-SXM4-40GB & Glow & 32 & 06:56:30 & 0.84 & 1.09e-07 & 1.25 & 0.79 & 1.97e-09\\
512 & 4 x A100-SXM4-40GB & Glow 8 blocks & 12 & 5:37:18 & 0.89 & 5.28e-08 & 1.01 & 0.82 & 1.43e-09\\
512 & 4 x A100-SXM4-40GB & WaveletFlow & 32 & 1:30:30 & 0.89 & 3.97e-08 & 3.68 & 0.83 & 7.83e-10\\
1024 & 4 x A100-SXM4-40GB & Glow 8 blocks & 4 & 6:26:27 & 0.89 & 5.35e-08 & 1.16 & 0.77 & 2.55e-09\\
1024 & 4 x A100-SXM4-40GB & WaveletFlow & 32 & 2:26:44 & 0.89 & 4.12e-08 & 2.27 & 0.81 & 2.06e-10\\
2048 & 4 x A100-SXM4-40GB & WaveletFlow & 32 & 6:00:56 & 0.90 & 4.09e-08 & 0.92 & 0.87 & 1.69e-10 \\
4096 & 4 x A100-SXM4-40GB & WaveletFlow & 16 & 7:10:48 & 0.88 & 5.43e-08 & 0.56 & 0.81 & 2.18e-10\\
\end{tabular}
\end{table*}

\subsection{Data-intensive regime scaling}
\label{sec:datascaling}

As previously stated, the main bottleneck of the forward modelling algorithm to
scale to larger datasets is consistent modelling of the visibility data which
corresponds to the $\mathcal{L}(\mathbf{x})$ term in the \eqref{eq:opt}.
However, one may rewrite the term in a form of expectation up to an additive
constant:
\begin{equation}
\begin{aligned}
\mathcal{L}(\mathbf{x}) &= \sum_{i}^N
  \frac{\left|\tilde{V}_i - V_i \right|^2}
{2 \sigma^2_i} = N \mathbb{E}_{i \sim \mathcal{U}\{1, N\}}
\left[
\frac{\left|\tilde{V}_i - V_i \right|^2}
{2 \sigma^2_i}
\right] \equiv \\
&\equiv N \mathbb{E}_{i \sim \mathcal{U}\{1, N\}}
\left[
  l(\mathbf{x})
\right].
\end{aligned}
\end{equation}
Therefore, the optimization problem can be rewritten in the following form:
\begin{equation}
\begin{aligned}
\mathbf{\theta} \longleftarrow& \argminB_{\mathbf{\theta}}
  \mathbb{E}_{z \sim \mathcal{N}(0,\, 1),\:i \sim \mathcal{U}\{1, N\} } \left[
    N l_i(\mathbf{x}) - \log \left|
        \det \frac{
          \partial \mathbf{x}
        }{\partial \mathbf{z}}
      \right| + \right. \\ &+ \left. \sum_i \lambda_i
      \mathcal{R}_{i}(\mathbf{x})
  \right] \equiv \argminB_{\mathbf{\theta}}
    \mathbb{E}_{z \sim \mathcal{N}(0,\, 1),\:i \sim \mathcal{U}\{1, N\} }
    \left[q_i(\mathbf{z})\right].
\end{aligned}
\end{equation}
This enables us to model only some fraction of visibility data on each
iteration, while maintaining the same stochastic optimization framework.

To test the approach, the algorithm was applied to the model observations using
both the HL Tau dataset and 'unaveraged' RadioAstron dataset UV-coverage and
errors. Each array configuration is tested for both model source maps. The runs
results are summarized in Table~\ref{tab:batching}. The term 'epochs'
denote the expected number of times the algorithm encounter each sample in the
dataset. Notably, considerable fidelity in the recovered images can be obtained
by the model while processing only $\approx 1\%$ of the HL Tau dataset.
However, smaller batch sizes require increasing the AdamW $\beta_1$ parameter
from its default value of 0.9 to 0.99.

\begin{table}%
\caption{Reconstruction fidelity in the data-intensive regime.
\label{tab:batching}}
\begin{tabular}{lrrrlrr}
        & SSIM & MSE & $\beta_1$ & model & batch &  epochs  \\
\hline
HL Tau  & 0.82 & 3.23e-09 & 0.9 & Disk & 320000 &  148.213644 \\
        & 0.81 & 3.25e-09 & 0.9 & Disk & 32000 &  14.821364 \\
        & 0.82 & 1.78e-09 & 0.99 & Disk & 32000 &  14.821364 \\
        & 0.79 & 3.26e-09 & 0.99 & Disk & 3200 & 1.482136 \\
        & 0.77 & 3.53e-09 & 0.99 & Disk & 320 & 0.148214 \\
        & 0.75 & 2.07e-09 & 0.99 & Disk & 32 & 0.014821 \\
        & 0.91 & 5.20e-08 & 0.9 & VLBI & 320000 & 148.213644 \\
        & 0.91 & 5.25e-08 & 0.9 & VLBI & 32000 &  14.821364 \\
        & 0.87 & 8.52e-08 & 0.99 & VLBI & 3200 &  1.482136 \\
        & 0.85 & 1.05e-07 & 0.99 & VLBI & 320 & 0.148214 \\
        & 0.83 & 9.02e-08 & 0.99 & VLBI & 32 & 0.014821 \\
\hline
M87*    & 0.91 & 9.10e-11 & 0.99 & Disk & 3200000 & 534.768562 \\
        & 0.93 & 2.18e-08 & 0.99 & VLBI & 3200000 & 534.768562 \\
\end{tabular}
\end{table}

\section{Conclusions and discussion}
\label{sec:disc}

In this work, we apply the normalizing flows approach to the problem of
interferometric  imaging. By introducing modified WaveletFlow architecture the
method requires less computational resources for sampling high-dimensional
image posteriors, compared to classical Bayesian sampling methods. On model
data, we demonstrate that prior assumptions of the algorithm enable reliable
imaging without being unnecessarily restrictive and biased towards expected
features of astrophysical sources, such as jets or rings, while maintaining
comparable fidelity of images to the classical algorithms results. We also
demonstrate that the method can be applied to both calibrated SVLBI
observational data, which tend to have a lower signal-to-noise ratio due to
limitations of space telescopes and larger baselines, and 'data-intensive'
interferometry, which effectively restricts the amount of full dataset
evaluations. In contrast with other machine learning approaches to imaging
\citep[see for example][]{Wang2023}, the algorithm remains relatively tractable
and does not require carefully prepared training sets of images. This enables
the user to consider it as an extension of classical maximum entropy methods.

The original work \citep{Sun2020} used closure quantities to take multiplicative
visibilities noise \eqref{eq:vis_errors} due to calibration errors into
account. However, this approach introduces several issues. The method requires
an analytical probability density function $p(y | x)$ in \eqref{eq:kldiv}.
In the first approximation, one can assume that the errors in closure
quantities follow a Gaussian distribution. However, in case of large baselines,
thermal noise significantly affects the visibility phases so closure quantities
are no longer invariant to calibration and Gaussian approximation no longer
works. Also, closure quantities are not independent, which is the case for
visibilities. Papers, addressing the approximated analytical form of the PDF
do exist \citep{Blackburn2020}, but one still needs to a find non-redundant set
of stations, which can be problematic in case of archive averaged data that
typically lacks some baselines due to low SNR. Future work will explore
alternative treatments of calibration errors, including joint modelling of
images and station gains, somewhat resembling self-calibration algorithms.

Beyond calibration, the reconstruction remains dependent on the choice of
regularizers and their corresponding weights. Selecting these parameters
typically involves manual or semi-automated surveys, as different weight
configurations can significantly alter the resulting image structures. While
existing strategies, such as multi-objective evolution
\citep{Muller2023_MO} or nested sampling \citep{Tiede2022}, attempt to map the
range of plausible solutions, they are often computationally prohibitive.

\section*{Acknowledgements}

The RadioAstron project was led at the Astrospace Center of the Lebedev Physics
Institute, Russian Academy of Sciences. and the Lavochkin Scientific and
Production Association under a contract with the State Space Corporation
ROSCOSMOS together in collaboration with partner organizations in Russia and
other countries.

This paper makes use of the following ALMA data: ADS/JAO.ALMA\#2011.0.00015.SV.
ALMA is a partnership of ESO (representing its member states), NSF (USA) and
NINS (Japan), together with NRC (Canada) and NSC and ASIAA (Taiwan), and KASI
(Republic of Korea), in cooperation with the Republic of Chile. The Joint ALMA
Observatory is operated by ESO, AUI/NRAO and NAOJ.

\section*{Data Availability}

The datasets and imaging maps used in this study are available from the
corresponding author upon reasonable request.

\bibliographystyle{mnras}

\appendix

\section{Regularizations}
\label{sec:apreg}

\begin{figure}
\begin{minipage}[h]{0.79\linewidth}
\center{\includegraphics[width=1\linewidth]{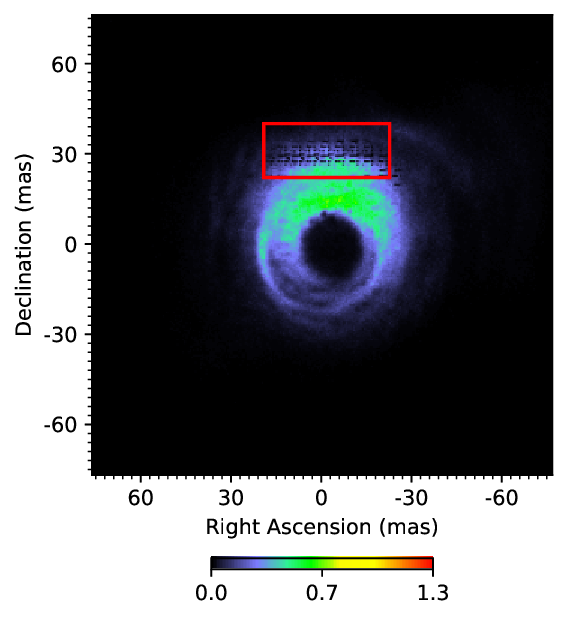}} \\
\end{minipage}
\caption{The example of the imaging artefacts: the 'grid' on the image
(highlighted by the red box).}
\label{fig:imagingartifacts}
\end{figure}

As was previously discussed, the regularizer term is considered to be a
logarithm of a prior ($\log p(x)$). We mostly follow the
\citet{Sun2020, Chael2018} approach and the regularizer has the following
form:
\begin{equation}
\begin{aligned}
\mathcal{R}(\mathbf{x})
  &= \lambda_{1} \text{loss\_ce}(\mathbf{x})
  + \lambda_{2} \text{loss\_TSV}(\mathbf{x})
  + \lambda_{3} \text{loss\_flux}(\mathbf{x}) \\
  &+ \lambda_{4} \text{loss\_l1\_dwt}(\mathbf{x})
\end{aligned}
\end{equation}
The main difference from \citet{Chael2018} is that closure quantities are
invariant to image translations, which is not the case for 'convex'
imaging so the specific loss term to centre the image is no longer required.
Let us define the remaining regularizer terms. The first term is
'entropy loss':
\begin{equation}
  \text{loss\_ce}(\mathbf{x}) = \sum_{m,n} I_{m, n} \left(
    \log(I_{m, n}) - \log(P_{m,n}) \right),
\end{equation}
where $P_{m,n}$ is a prior image and usually set to a Gaussian function with
a desired flux $F$. This regularizer follows \citet{Narayan1986} intuition and
results in 'less commitment' image. The second term forces an image to be
smooth:
\begin{equation}
  \text{loss\_TSV}(\mathbf{x}) = \sum_{m,n}
  (I_{m + 1,n} - I_{m,n})^2 + (I_{m,n + 1} - I_{m,n})^2.
\end{equation}
The third term forces the correct image flux and it is somewhat
redundant if zero-length baselines present in the dataset:
\begin{equation}
  \text{loss\_flux}(\mathbf{x}) = \left(\sum_{m, n} I_{m,n} - F\right)^2.
\end{equation}
And finally, the last term applies l1 regularization on the wavelet coefficients
sampled from the model (before the exponentiation), forcing the image to be
compressible. The choice of the regularizers coefficient influence the
recovered image significantly introducing imaging artefacts. The example is
shown in Fig.~\ref{fig:imagingartifacts}.

\section{Convolutional neural network architecture}
\label{sec:apconv}

\begin{figure}
\includegraphics[scale=0.21]{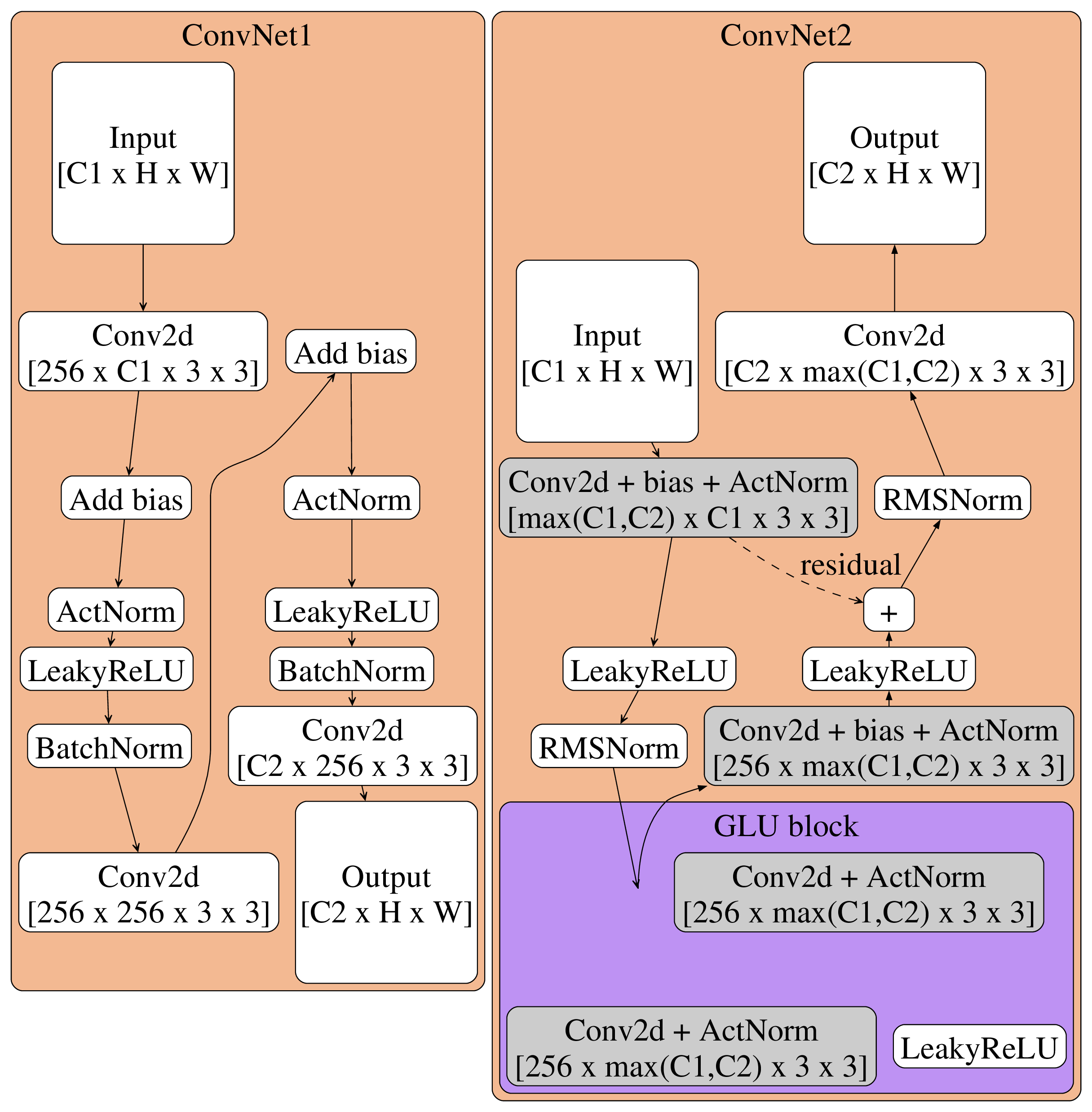}
\caption{ Two convolutional architectures used in the coupling layers.
(Left) ConvNet1: the architecture adapted from \citet{Sun2020} source code
repository.
(Right) ConvNet2: residual architecture addressing the large memory throughput
requirements of the first architecture.
\label{fig:archconvnet}}
\end{figure}

Two convolutional architectures were tested. The first ('ConvNet1' in Fig.
~\ref{fig:archconvnet}) was motivated by a similar architecture present in
the \citet{Sun2020} source code repository. It was found that the second
convolutional layer was the most computationally expensive operation. This was
addressed by using 16-bit precision float values (as this part of the network
is not required to be invertible, but only to be deterministic) and by
introducing a residual architecture ('ConvNet2' in Fig.~\ref{fig:archconvnet}).
To minimize the required throughput, biases were removed wherever possible and
actnorm layers were fused into convolutional kernels. RMSNorm was also used
instead of BatchNorm, as it enables the use of single-image batches during
training. This is required for techniques addressing low-memory conditions,
such as gradient accumulation.

\bsp	%
\label{lastpage}
\end{document}